\documentclass[twocolumn]{aastex62}
\usepackage{comment}
\pdfoutput=1
\newcommand{\htwo}{H\,{\scriptsize\sc{II}} }

\begin{document}
\title{The Infall Motion in the Low-Mass Protostellar Binary NGC 1333 IRAS 4A1/4A2}

\author{Yu-Nung Su}
\affiliation{Institute of Astronomy and Astrophysics, Academia Sinica, 11F of ASMAB, AS/NTU No.1, Sec. 4, Roosevelt Road, Taipei 10617, Taiwan}

\author{Sheng-Yuan Liu}
\affiliation{Institute of Astronomy and Astrophysics, Academia Sinica, 11F of ASMAB, AS/NTU No.1, Sec. 4, Roosevelt Road, Taipei 10617, Taiwan}

\author{Zhi-Yun Li}
\affiliation{Department of Astronomy, University of Virginia, Charlottesville, VA 22904, USA}

\author{Chin-Fei Lee}
\affiliation{Institute of Astronomy and Astrophysics, Academia Sinica, 11F of ASMAB, AS/NTU No.1, Sec. 4, Roosevelt Road, Taipei 10617, Taiwan}

\author{Naomi Hirano}
\affiliation{Institute of Astronomy and Astrophysics, Academia Sinica, 11F of ASMAB, AS/NTU No.1, Sec. 4, Roosevelt Road, Taipei 10617, Taiwan}

\author{Shigehisa Takakuwa}
\affiliation{Institute of Astronomy and Astrophysics, Academia Sinica, 11F of ASMAB, AS/NTU No.1, Sec. 4, Roosevelt Road, Taipei 10617, Taiwan}
\affiliation{Department of Physics and Astronomy, Graduate School of Science and Engineering, Kagoshima University, 1-21-35 Korimoto, Kagoshima 890-0065, Japan}

\author{I-Ta Hsieh}
\affiliation{Institute of Astronomy and Astrophysics, Academia Sinica, 11F of ASMAB, AS/NTU No.1, Sec. 4, Roosevelt Road, Taipei 10617, Taiwan}

\begin{abstract}

 We report ALMA observations of NGC 1333 IRAS 4A, a young low-mass protostellar binary, referred as 4A1 and 4A2. With multiple H$_2$CO transitions and HNC (4$-$3) observed at a resolution of 0$\farcs$25 ($\sim$70 au), we investigate the gas kinematics of 4A1 and 4A2. Our results show that, on the large angular scale ($\sim$10$\arcsec$), 4A1 and 4A2 each display a well-collimated outflow along the N-S direction, and an S-shaped morphology is discerned in the outflow powered by 4A2. On the small scale ($\sim$0$\farcs$3), 4A1 and 4A2 exhibit distinct spectral features toward the continuum centroid, with 4A1 showing simple symmetric profiles predominantly in absorption and 4A2 demonstrating rather complicated profiles in emission as well as in absorption. Based on radiative transfer modeling exercises, we find that the physical parameters inferred from earlier low-resolution observations cannot be directly extrapolated down to the 4A1 inner region. Possible reasons for the discrepancies between the observed and modelled profiles are discussed. 
 We constrain the mass infall rate in 4A1 to be at most around 3$\times$10$^{-5}$ \emph{M}$_{\odot}$ year$^{-1}$ at the layer of 75 au. For the kinematics of the 4A2 inner envelope, the absorbing dips in the H$_2$CO spectra are skewed toward the redshifted side and likely signatures of inward motion. These absorbing dips are relatively narrow. This is, like the case for 4A1, significantly slower than the anticipated inflow speed. We estimate a mass infall rate of 3.1$-$6.2 $\times$ 10$^{-5}$ \emph{M}$_\sun$ year$^{-1}$ at the layer of 100 au in 4A2. 

\end{abstract}

\section{Introduction \label{s-intro}}
Although gravitational collapse of molecular gas has been widely accepted from the theoretical viewpoints of star formation, acquiring direct observational evidence of gas infall motions poses a challenge \citep[][]{mye00,eva15}. Based on single-dish observations, infall motions are generally inferred from asymmetric blue profiles of optically thick lines \citep[e.g.,][]{mar97}. Such kinematic features at relatively low velocity, however, are easily contaminated by other star formation activities such as outflows, rotation, and turbulence. Another diagnostic of infall motion is the triangular-like position-velocity (PV) structure in both blue- and red-shifted sides \citep{oha97}. Such types of PV structure have been reported in several low-mass star-forming regions (e.g., HH~211, HH~212, and L1527) and is interpreted as the result of an infalling-rotating envelope \citep{sak14,lee17,lee19}. 

The ``inverse P-Cygni'' line profile, i.e., the redshifted self-absorption of spectral lines, is considered as a more robust probe of infall motions. Envelope material in front moving inward toward the central (continuum) source results in the redshifted absorption. The absorption signature is easier to discern with observations at higher frequencies because the intensity of the dust continuum increases rapidly with frequency. Indeed, high frequency ($\sim$1 THz) observations with \emph{SOFIA} and \emph{Herschel} have identified redshifted absorption features toward several star formation regions \citep{wyr12,mot13}. With angular resolutions of 15\arcsec$-$30\arcsec, however, these studies can not resolve the kinematics in the inner infalling envelope. For free-fall inward motion (i.e., $\emph{v}_{infall}$ $\propto$ \emph{r}$^{-1/2}$), one would expect a higher infalling velocity closer to the central object. Such high-velocity infalling gas can not be readily discerned with single-dish observations due to its relatively small filling factor.  The fact that the dust continuum could become optically thick at high-frequency regimes, in particular in the inner region, will further complicate the detection of high-velocity infalling gas. Interferometric observations at millimeter and sub-millimeter wavelengths serve as another channel to probe infalling gas \citep[e.g.,][]{dif01}. With high sensitivity and angular resolution, ALMA provides an invaluable opportunity to study inward motion in molecular lines with their absorption signature against the continuum.  For example, recent ALMA 690 GHz observations of IRAS 16293$-$2422B at 0\farcs2~resolution revealed pronounced inverse P-Cygni profiles, with faster infalling gas located closer to the central object \citep{zap13}. Observations of the low-mass protostar B335 with ALMA at 0\farcs5~resolution also identified redshifted absorption features against the continuum in HCN and HCO$^{+}$ lines \citep{eva15}. 

Located at a distance of 293 pc in the Perseus star formation complex \citep{ort18,zuc18}, NGC 1333 IRAS 4A (hereafter IRAS 4A)  is one of the brightest dust continuum sources among low-mass protostars \citep{loo00}. Single dish observations of IRAS 4A with millimeter and submillimeter transitions revealed asymmetric blue profiles of optically thick lines relative to symmetric optically thin transitions, suggestive of infall motions in this region \citep[e.g.,][]{bla95, bel06}.  The inverse P-Cygni profiles toward IRAS 4A were obtained in H$_2$CO (3$_{12}-2_{11}$) as well as CS (3$-$2) and N$_2$H$^{+}$ (1$-$0) at $\sim$2$\arcsec$ resolution with the PdBI \citep{dif01}. This detection provided the least ambiguous evidence of infall motions in this region to date, although the possibility of the absorption features being caused by an unrelated foreground absorption layer can not be completely ruled out \citep[for a comprehensive discussion, see][]{cho04}.  Observations of polarized dust emission with the SMA have revealed a clear hourglass shaped B-field toward IRAS 4A  \citep{gir06}. The estimated mass-to-magnetic flux ratio implies that the region traced by SMA is slightly supercritical. \citet{gon08} modeled the observed magnetic field in this region and concluded that the magnetic field morphology in IRAS 4A is consistent with the standard theoretical scenario for the formation of low-mass stars from cores threaded by ordered rather than turbulent magnetic fields.

IRAS 4A has been resolved into two separated sources (referred as 4A1 and 4A2, with an angular separation of 1\farcs8) at high-resolution mm/submm maps \citep[e.g.,][]{loo00, jor07, che13}.  The two cores, 4A1 and 4A2, each drive a well-collimated outflow/jet along the N-S direction \citep{cho05,san15}. VLA observations of NH$_3$ (2,2) and (3,3) lines revealed a rotating disk toward 4A2 only \citep{cho10}. In contrast, the nature of NH$_3$ lines toward 4A1 is less clear. It could be a mixture of a circumstellar disk and a molecular outflow \citep{cho11}. With high-resolution 8 mm VLA dust continuum observations, the recent VANDAM survey produced disk-fit results of 4A1 in the visibility domain \citep{seg18}. As signposts of hot corinos, complex molecules (e.g., HCOOCH$_3$, HCOOH, and CH$_3$CN) have been detected toward IRAS 4A by the IRAM 30-m Telescope \citep{bot04}. Recent interferometric observations with ALMA and PdBI further resolved the distributions of these complex species, with 4A2 demonstrating a chemically rich spectrum and 4A1 lacking detections of most complex organic molecules \citep{taq15,san15,lop17}. Based on the different properties of the molecular outflows and the central driving sources, \citet{san15} argued  different evolutionary stages for 4A1 and 4A2, with 4A1 being younger than 4A2. A similar conclusion was also reached by \cite{cho10} with the VLA observations in NH$_3$ and SiO.

Here we present newly obtained ALMA observations in multiple H$_2$CO transitions as well as HNC (4$-$3) at a resolution of 0\farcs25 (corresponding to linear scale of $\sim$70 au at a distance of 293 pc) to illustrate the gas kinematics around the two dense cores 4A1 and 4A2 on scales of 70 au -- 350 au.  The obtained ALMA datasets also contain rich information in chemistry, and the chemical behaviors of 4A1 and 4A2 have been discussed in \cite{dip19}, with their main finding in the firm identification, for the first time, of complex organic molecules (COMs), including CH$_3$OH, $^{13}$CH$_3$OH, CH$_2$DOH, and CH$_3$CHO, associated with the 4A1 core in absorption. Our ALMA observations provide to date the data with the best spatial resolution to discriminate the gas kinematics between 4A1 and 4A2. We note that the H$_2$CO and HNC data presented here have a velocity resolution of $\sim$0.12 km s$^{-1}$, about a factor of 8 better than the data presented by \citet{dip19}.

\begin{deluxetable*}{lcllcl}
\singlespace
 \tablecaption{\sc Expected Transitions \label{tb-line-list}}
 \tablewidth{0pt}
 \tablehead{\multicolumn{3}{c}{Lines in LSB} & \multicolumn{3}{c}{Lines in USB} \\
                   \colhead{Transition} & \colhead{Frequency}  & \colhead{Energy level} & \colhead{Transition} & \colhead{Frequency}  & \colhead{Energy level} \\
            \colhead{} & \colhead{(GHz)} & \colhead{(K)} & \colhead{} & \colhead{(GHz)} & \colhead{(K)} }
 \startdata
        H$_2$CO (5$_{15}$$-$4$_{14}$)	 & 351.76864	 &  E$_u$=62.5 K & H$_2$CO (5$_{05}$$-$4$_{04}$) 	& 362.73605 	& E$_u$=52.3 K  \\
            					                           &        		     &			              & H$_2$CO (5$_{24}$$-$4$_{23}$) 	& 363.94589 	& E$_u$=99.5 K  \\
                             	 			              &        		     &           		      & H$_2$CO (5$_{41}$$-$4$_{40}$) 	& 364.10325 	& E$_u$=240.7 K  \\
                              				              &              	 &           		      & H$_2$CO (5$_{33}$$-$4$_{32}$) 	& 364.27514 	& E$_u$=158.4 K  \\
				                                        &        		     &           		      & H$_2$CO (5$_{32}$$-$4$_{31}$) 	& 364.28888 	& E$_u$=158.4 K  \\
                              				              &        		     &           		      & HNC (4$-$3) 			 & 362.63030 	& E$_u$=43.5 K  \\
\enddata
\end{deluxetable*}

\section{Observations and Data Reduction}
Observations of NGC 1333 IRAS 4A with ALMA in band 7 (i.e., the 350 GHz band) were carried out using its 12-m Array in its C40$-$3 and C40$-$5 configurations under the project code 2015.1.00147.S.  For the observations in the C40$-$5 configuration, two executions were carried out on 2016 July 23 and 24, with 39 and 42 antennas in the array, respectively. Each execution has an on-source time of about 28 minutes, and the projected baseline lengths range from 15.4 m to 1100 m. For the observations in the C40$-$3 configuration, a single execution was obtained on 2016 December 14, with 43 antennas in the array. The on-source integration time of the C40-3 execution is also about 28 mins, and the projected baseline lengths range from 12.3 m to 382 m. The phase center in the ICRS reference frame was R.A. = 03$^h$29$^m$10.50$^s$ and decl. = $+$31\degr13\arcmin31\farcs50 (J2000).  The half-power width of the ALMA 12-m antenna primary beam was $\sim$16\farcs55 at 351.7 GHz, more than sufficient to cover the entire extents of the 4A1 and 4A2 cores.

The ALMA band 7 receivers were used to capture simultaneously six H$_2$CO transitions with their upper energy levels ranging from approximately 50 K to 240 K and HNC (4$-$3), as summarized in Table~\ref{tb-line-list}.  For each transition, the ALMA correlator was configured to provide a velocity resolution of about 0.12 km s$^{-1}$ and a total bandwidth of either 58.6 MHz ($\sim$ 50 km s$^{-1}$) or 117.2 MHz ($\sim$100 km s$^{-1}$). A broadband spectral window of 1.875 GHz centered at 350.7 GHz with a resolution of 1.13 MHz (0.965 km s$^{-1}$) was also included, and the results of the broadband window data have been presented in \cite{dip19}. Quasars J0238+1636 and J0510+1800 were observed to calibrate the passband, and Ceres and quasars J0006-0623 and J0510+1800 were observed to calibrate the absolute flux scale. The quasar J0336+3218 was observed as the gain calibrator.  The acquired visibility data were reduced using the observatory pipeline in CASA (Common Astronomy Software Application; McMullin et al. 2007) version 4.7.0. We then generated both the 0.84~mm continuum and spectral visibilities by fitting and subtracting continuum emission in the visibility domain. The calibrated visibility data were Fourier-transformed and CLEANed with CASA 4.7.0 also. By using Briggs weighting with a robust parameter of 0.5, we obtained a synthesized beam size about 0\farcs31 $\times$ 0\farcs20 at P.A. of $-$6.0$^\circ$. We smoothed our data to 0.15 km s$^{-1}$ resolution for the analysis presented below. The typical rms noise level in the maps is about 6$-$8 mJy beam$^{-1}$ ($\sim$0.95$-$1.20 K) in a 0.15 km s$^{-1}$ bin. We note that given the existence of the absorption features (see \S\ref{result-core}), a preferred approach of a CLEAN algorithm is to perform continuum subtraction after CLEANing in order to avoid using CLEAN on a cube, where both emission and absorption are present. The images made with these two approaches, however, do not exhibit notable difference in line features. In particular, the shapes of the spectral profiles toward 4A1 and 4A2 made with these two CLEAN approaches are almost identical.

\begin{figure}
 \vspace{-1.4cm}
 \hspace{-2.9cm} 
 \includegraphics[width=4.4in,angle=270]{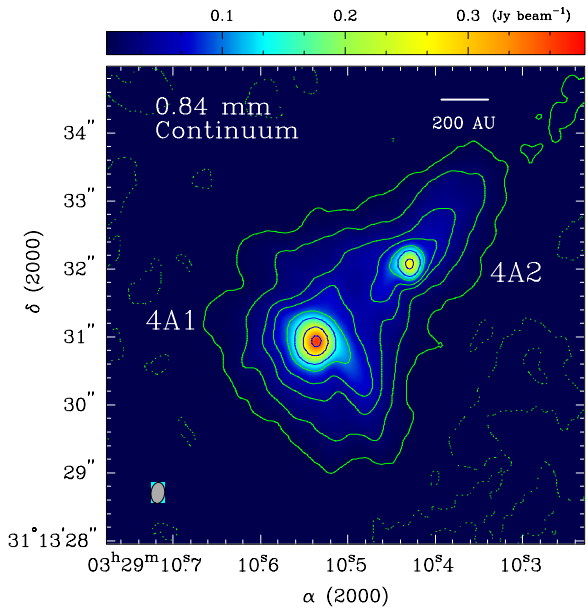}
 \vspace{-1.6cm}
 \caption{0.84~mm continuum emission of IRAS 4A. Contour levels are $-$10, $-$3, 3, 10, 20, 30, 50, 75, 125, 175 $\times$ 2 mJy beam$^{-1}$. The dark ellipse at the bottom left denotes the synthesized beam of 0\farcs31 $\times$ 0\farcs20 at P.A. $-$6.45$^\circ$. Note that due to the existence of bright sources, the continuum map is dynamic range limited rather than sensitivity limited.\label{cont}}
\end{figure}

\section{The 0.84~mm Continuum Emission \label{resultcont}}
Figure~\ref{cont} shows the 0.84~mm (357 GHz) continuum image toward IRAS 4A. At an angular resolution of 0\farcs31 $\times$ 0\farcs20, IRAS 4A is clearly resolved into two components, with 4A1 located in the SE and  4A2 in the NW.  The brightness temperatures inferred from the observed peak continuum are 57 K and 42 K toward 4A1 and 4A2, respectively. Note that these brightness temperatures are the beam-averaged values, and the actual peak brightness temperatures at smaller scales could be higher. 4A1 and 4A2 each appear to consist of a compact core and an extended outer envelope. The results of the 2D Gaussian fits as well as the estimated mass of each component can be found from Table~1 of \cite{dip19}. With the optically-thin assumption, the authors estimated the dust and gas mass to be 0.15 \emph{M}$_\sun$, 0.49 \emph{M}$_\sun$, 0.07 \emph{M}$_\sun$, and 0.50 \emph{M}$_\sun$ for the 4A1 compact component, 4A1 extended component, 4A2 compact component, and 4A2 extended component, respectively. As  concluded by \cite{dip19}, the 0.84~mm dust emission from the 4A1 core is optically thick and 4A2 is predominantly optically thin. As a consequence, the estimated mass for the 4A1 compact component could be significantly underestimated. \cite{dip19} also derived the molecular hydrogen column density toward the 4A1 centroid to be at least 1.3 $\times$ 10$^{26}$ cm$^{-2}$. We note that in addition to the above-mentioned compact and extended components, several spur-like structures can be discerned in the 0.84 mm continuum emission. The most evident one is that located to the south-west of 4A1.

\begin{figure*}
\vspace{-2.2cm} 
\hspace{-.9cm} 
 \includegraphics[width=6.1in,angle=270]{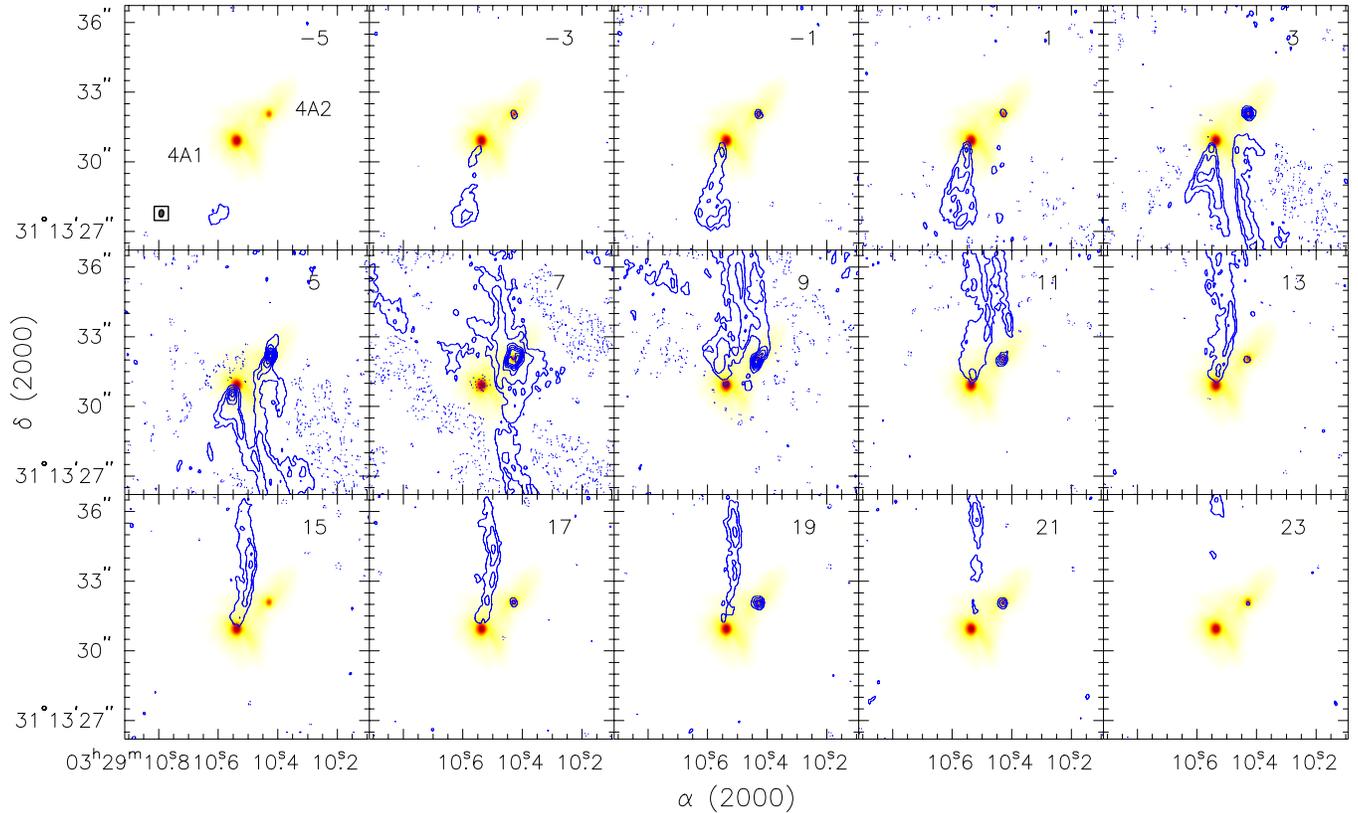}
 \vspace{-2.3cm} 
 \caption{Channel maps of IRAS 4A in H$_2$CO 5$_{05}-4_{04}$. The local standard of rest (LSR) velocity of each channel is indicated in the upper right corner of each panel. The LSR velocity of the system is 6.95 km s$^{-1}$. Solid contours are at 4, 10, 20, 30, 40, 50, 60 and 70 $\times$ 3 mJy beam$^{-1}$, and dotted contours indicate $-$4, $-$10, $-$20, and $-$30 $\times$ 3 mJy beam$^{-1}$. For each panel, the overlaid color scales are the 0.84 mm continuum emission shown in Figure~\ref{cont}. The dark ellipse at the bottom left of the first panel denotes the synthesized beam. The compact component located at the 4A2 continuum peak detected from 17 km s$^{-1}$ to 23 km s$^{-1}$ is likely not related to the H$_2$CO 5$_{05}-4_{04}$ outflowing gas but comes from contamination of other molecular species since 4A2 is a known ``hot corino" object. \label{outflow-chan}} 
\end{figure*}

\begin{figure*}
\vspace{-5.4cm} 
\hspace{-1.2cm} 
\includegraphics[width=6.2in,angle=270]{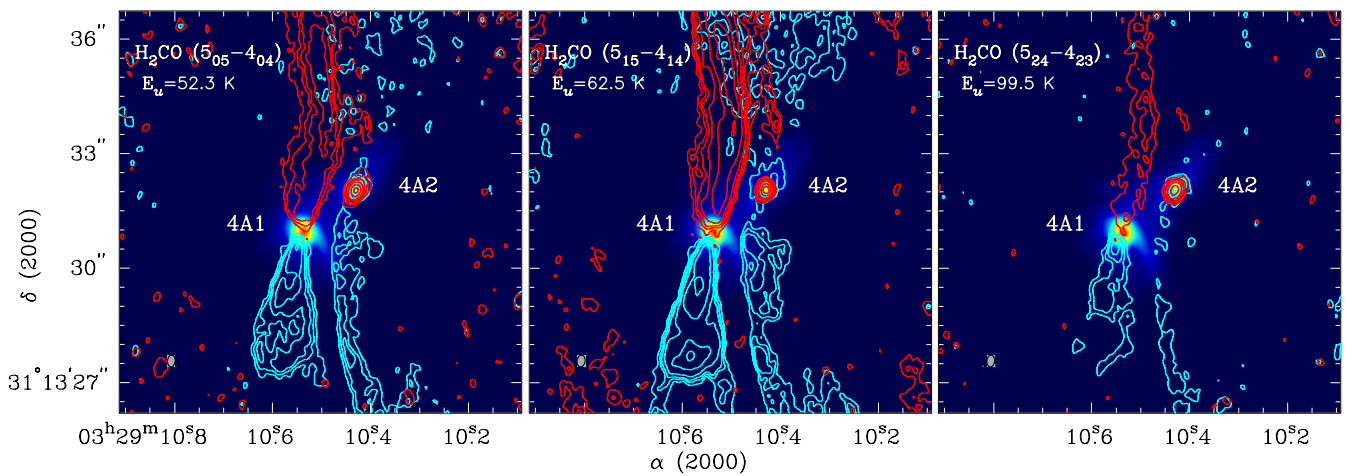}
\vspace{-4.2cm} 
\caption{Bipolar molecular outflows of H$_2$CO 5$_{05}$$-$4$_{04}$ (\emph{left}), 5$_{15}$$-$4$_{14}$ (\emph{middle}) and 5$_{24}$ $-$4$_{23}$ (\emph{right}) toward IRAS 4A. In each panel, the color scales present the 0.84 mm continuum shown in Figure~\ref{cont} and the dark ellipse at the bottom left denotes the synthesized beam. \label{outflow-int}} 
\end{figure*}

\section{H$_2$CO and HNC (4$-$3) Features \label{resultline}}
The spectral datacubes of the H$_2$CO and HNC (4$-$3) lines exhibit both compact features associated with 4A1 and 4A2 and extended structures in the vicinity of IRAS 4A. Determining the systemic velocities of both 4A1 and 4A2 is an essential step for a proper understanding of the observed kinematic features. \citet{dif01} estimated a systemic velocity of 6.95 km s$^{-1}$ for the IRAS 4A system as a whole. 4A1 and 4A2, however, may have (slightly) different systemic velocities. By using the average fitted peak velocity of the six observed H$_2$CO transitions (see Table \ref{tb-line-fit} as well as \S\ref{res_4a1} and \S\ref{res_4a2} also), we adopt systemic velocities of 6.86 km s$^{-1}$ and 7.00 km s$^{-1}$, respectively, for 4A1 and 4A2. Note that the blue- and red-shifted motions mentioned in the following sections are relative to the above-mentioned systemic velocities. We present below the results of the spectral line features from the large angular scale to the small angular scale.

\subsection{Molecular Outflows \label{resultoutflow}}
The most noticeable extended structures are the two collimated molecular outflows, which can be discerned in all the observed H$_2$CO transitions as well as in HNC (4$-$3), powered by 4A1 and 4A2. Figure \ref{outflow-chan} shows the channel maps of the H$_2$CO 5$_{05}-4_{04}$ line toward IRAS 4A. For display purposes, we have smoothed the channel maps to a velocity resolution of 2 km s$^{-1}$. Figure \ref{outflow-int} displays the integrated blueshifted and redshifted emission of H$_2$CO 5$_{05}-4_{04}$, 5$_{24}-4_{23}$, and 5$_{15}-4_{14}$. For both Figures, two well-collimated outflows along the N-S direction at $\sim$8$\arcsec$ scale are visible. The 4A1 outflow seen in H$_2$CO 5$_{05}-4_{04}$ is detected to relative velocities of about 12 km s$^{-1}$ and 16 km s$^{-1}$, respectively, in its blue-shifted and red-shifted components. In contrast, the H$_2$CO 5$_{05}-4_{04}$ outflowing gas powered by 4A2 is only detected up to $\sim$5 km s$^{-1}$ in relative velocity. Note that the lack of emission features toward the centroid of 4A1 in Figures \ref{outflow-chan} and \ref{outflow-int} also imply its 0.84~mm continuum is optically thick.

The low-velocity blue-shifted H$_2$CO component associated with 4A1 exhibits a conical shell-like structure opening to the south-south-east. In contrast, the high-velocity blue-shifted component shows a jet-like structure.  Such distinct morphologies in low- and high-velocity components have been detected in several low-mass outflows such as those from HH211 and IRAS 04166+2706 \citep{gue99, san09}. Regarding the red-shifted component of the 4A1 outflow, no clear conical structure can be discerned, perhaps due to contamination or overlap with the 4A2 outflow. Our observations also detect in H$_2$CO 5$_{15}$$-$4$_{14}$ and  5$_{33}$$-$4$_{32}$ emission the two extremely high-velocity (EHV) redshifted knots 2\arcsec$-$4\arcsec~from the core, with velocities (\emph{V}$_{\rm LSR}$) in the range of 35 $-$ 55 km s$^{-1}$, associated with 4A1 as previously reported by \citet{san15}. For the other observed H$_2$CO transitions as well as HNC (4$-$3), the spectral windows were not wide enough to capture the red-shifted EHV components.   

For the 4A2 H$_2$CO outflow, bending (or S-shaped) morphologies can be discerned in both the blue- and red-shifted components. The bending structure has been previously identified in CO (2$-$1) and SiO (5$-$4) with the PdBI observations \citep{san15}.  Within $\sim$8$\arcsec$ in radius of 4A2, \cite{san15} reported a collimated outflow along the N-S direction.  A sharp bend can be found at a distance of about 4$\arcsec$ away from the driving source, with the outflow direction changing to NW-SE. The authors interpreted such a sharp bend as resulting from outflow precession, and estimated jet precession on very short timescales of $\sim$200$-$600 years. Instead of a well-collimated morphology, our H$_2$CO maps further reveal a gradual change of the outflow direction in the close vicinity of 4A2.

\begin{figure*}
 \hspace{-1.1cm}
 \includegraphics[width=8.0in]{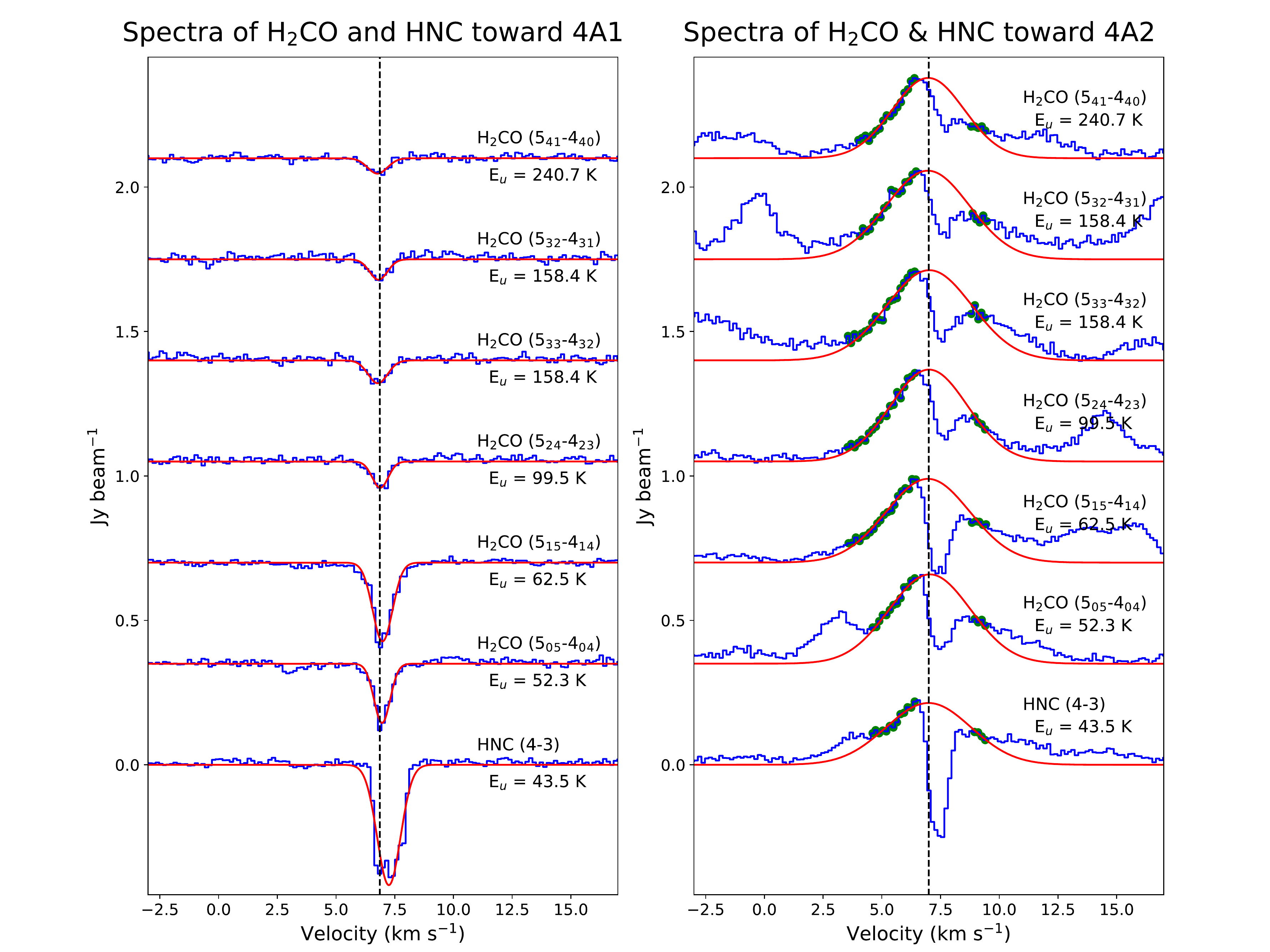}
 \caption{The continuum-subtracted spectra of the six H$_2$CO lines as well as HNC (4$-$3) toward the central positions of 4A1 (\emph{left}) and 4A2 (\emph{right}) in the sequence of excitation from bottom to top. The transitions and the corresponding upper-state excitation energies are also labelled. The fitted Gaussian profiles are overlaid. Note that in the case of 4A2, only partial data points (marked as green points) are used in the fitting. Data points which are contaminated by infall or outflow motions are excluded in the fitting. The dotted line denotes the systemic velocity of 4A1 (\emph{left}) and 4A2 (\emph{right}) at 6.86 km s$^{-1}$ and 7.00 km s$^{-1}$, respectively, as estimated from the average central line velocity of the six H$_2$CO lines. Note that the brightness temperatures inferred from the observed peak continuum are 57 K and 42 K toward 4A1 and 4A2, respectively (1 K equivalent to $\sim$7 mJy beam$^{-1}$). \label{H2CO_spec_fitting}}
\end{figure*}

\begin{deluxetable*}{llcccccc} 
\vspace{-.1cm}
\tabletypesize{\footnotesize}
\tablecaption{\sc Molecular transitions detected toward IRAS~4A1 and 4A2 \label{tb-line-fit}} 
\tablehead{\colhead {Transition}  &  \colhead {\emph{E}$_{u}$}  & \colhead {Peak Intensity} & \colhead {\emph{V}$_{\rm LSR}$} &\colhead{$\Delta$\emph{V}} & \colhead {Peak Intensity} & \colhead {\emph{V}$_{\rm LSR}$ } &\colhead{$\Delta$\emph{V}}\\ \colhead{} &\colhead{(K)}&\colhead{(Jy beam$^{-1}$)}&\colhead{(km s$^{-1}$)}&\colhead{(km s$^{-1}$)}&\colhead{(Jy beam$^{-1}$)}&\colhead{(km s$^{-1}$)}&\colhead{(km s$^{-1}$)} }
 \startdata
\colhead{} &\colhead{}&\multicolumn{3}{c}{transitions toward IRAS 4A1}& \multicolumn{3}{c}{transitions toward IRAS 4A2}\\
\hline
 HNC (4$-$3)                 	   &  43.5   & $-$0.417 (0.009) & 7.25 (0.01) & 1.16 (0.03) & 0.214 (0.009) & 6.99 (0.06) & 4.26 (0.19) \\  
 H$_2$CO (5$_{05}$$-$4$_{04}$)     &  52.3   & $-$0.205 (0.016) & 6.96 (0.03) & 0.81 (0.07) & 0.309 (0.008) & 7.05 (0.03) & 4.08 (0.10)\\  
 H$_2$CO (5$_{15}$$-$4$_{14}$)     &  62.5   & $-$0.273 (0.014) & 6.98 (0.02) & 0.99 (0.06) & 0.290 (0.007) & 6.96 (0.04) & 4.20 (0.10) \\  
 H$_2$CO (5$_{24}$$-$4$_{23}$)     &  99.5   & $-$0.091 (0.017) & 6.88 (0.07) & 0.80 (0.17) & 0.318 (0.007) & 7.01 (0.03) & 3.87 (0.07) \\  
 H$_2$CO (5$_{33}$$-$4$_{32}$)     &158.4    & $-$0.079 (0.015) & 6.78 (0.09) & 0.98 (0.22) & 0.312 (0.010) & 7.02 (0.05) & 4.29 (0.12) \\  
 H$_2$CO (5$_{32}$$-$4$_{31}$)     &158.4    & $-$0.071 (0.016) & 6.81 (0.10) & 0.90 (0.23) & 0.307 (0.008) & 6.96 (0.04) & 4.17 (0.11) \\  
 H$_2$CO (5$_{41}$$-$4$_{40}$)     & 240.7   & $-$0.053 (0.015) & 6.76 (0.14) & 0.99 (0.32) & 0.278 (0.010) & 6.98 (0.04) & 3.68 (0.11) \\  
\enddata
\end{deluxetable*}

\subsection{Dense Gas associated with Disk/Envelope} \label{result-core}
Figure~\ref{H2CO_spec_fitting} shows the observed spectra of the six H$_2$CO lines as well as HNC (4$-$3) at the pixel where the 0.84~mm continuum peaks at 4A1 (\emph{left}) and 4A2 (\emph{right}). Although toward both sources all targeted H$_2$CO and HNC transitions are clearly detected, 4A1 and 4A2 exhibit distinct spectral features, with 4A1 showing simple symmetric profiles predominantly in \emph{absorption} and 4A2 demonstrating relative complicated profiles in emission as well as in absorption.  

\subsubsection{The Line Feature toward 4A1  \label{res_4a1}}
The absorption profiles of the H$_2$CO and HNC transitions detected toward the 4A1 centroid appear to be Gaussian-like. We performed Gaussian profile fitting to extract the line parameters. The fitting profiles are overlaid in the \emph{left} panel of Figure~\ref{H2CO_spec_fitting}, and the fitting results are summarized in Table~\ref{tb-line-fit}. The 4A1 systemic velocity of 6.86 km s$^{-1}$ estimated from the average fitted peak velocity of the six observed H$_2$CO transitions is also labelled in Figure~\ref{H2CO_spec_fitting}. The line widths of the H$_2$CO transitions range from 0.80 km s$^{-1}$ to 0.99 km s$^{-1}$, and their central line velocities (\emph{V}$_{\rm LSR}$) are consistent with each other, i.e., in the range of 6.76 km s$^{-1}$ to 6.98 km s$^{-1}$. 

With a broader line-width of 1.16 km s$^{-1}$ and a redshifted central line velocity of 7.25 km s$^{-1}$, the absorption feature of HNC (4$-$3) is different from the above-mentioned H$_2$CO transitions. Unlike the compact ($<$1$\arcsec$) H$_2$CO absorption features detected toward 4A1 (see below), the absorption seen in HNC (4$-$3) is fairly extended. In particular, at velocities between 7.15 km s$^{-1}$ and 7.60 km s$^{-1}$, deep HNC (4$-$3) absorption can be discerned toward the whole extent of the 0.84~mm continuum, with almost all continuum emission absorbed. The extended HNC (4$-$3) absorption seen between 7.15 km s$^{-1}$ and 7.60 km s$^{-1}$ may result from outer envelope of 4A1 and 4A2 or even diffuse molecular gas located in NGC 1333 cloud.  Given that a velocity gradient is observed between 4A1 and 4A2 (see \S \ref{resultline}), we suspect at large scale a velocity gradient can be discerned too, as gas motions may take place from the inner to the outer envelope surrounding the IRAS 4A. The extended HNC (4$-$3) absorption, however, shows no sign of velocity gradient across the extent of the IRAS 4A continuum. Therefore, we favor the HNC (4$-$3) absorption observed between 7.15 km s$^{-1}$ and 7.60 km s$^{-1}$ tracing an extended redshifted foreground layer, although the possibility of a common envelope of 4A1 and 4A2 or even diffuse molecular gas in NGC 1333 can not be fully ruled out.

This foreground layer of material has previously been implied by \cite{lis88} in CO (1$-$0) and by \cite{yil13} in O$_2$~(3$_3-1_2$). 
With Herschel observations of multiple water lines, recently \cite{mot13} reported a low density foreground absorption layer at \emph{V}$_{\rm LSR}$ of 8.0 km s$^{-1}$, redshifted by about 1~km~s$^{-1}$ with respect to the systemic velocity of IRAS 4A. \cite{mot13} further argued that both the infall in the IRAS 4A envelope and the foreground layer are required in their model spectra calculations to reproduce well the observed water line profiles. Limited by coarse angular resolutions, the above observations unfortunately were not able to differentiate gas kinematics between 4A1 and 4A2 from the observed line profiles.
We conclude that HNC (4$-$3) absorption traces not only the dense gas immediately surrounding 4A1 but also this previous identified redshifted foreground material, resulting in the HNC (4$-$3) line width broader than that of the H$_2$CO lines. The absorption of the foreground layer also makes the HNC (4$-$3) central line velocity shift redward. This foreground layer could be part of a converging flow that is involved in the formation of the dense gas that collapsed into the IRAS 4A system or an independent gas layer. 

Comparing with the HNC (4$-$3) transition, the observed H$_2$CO transitions listed in Table~\ref{tb-line-list} all have higher energy levels and effective excitation density \citep{shi15}, and hence are unlikely to trace the foreground gas. For example, at a gas temperature of 20 K, the effective excitation density of HNC (4$-$3) is 2.4 $\times$ 10$^5$ cm$^{-3}$ only, about an order of magnitude smaller than that of H$_2$CO 5$_{05}-4_{04}$ (3.8 $\times$ 10$^6$ cm$^{-3}$) and H$_2$CO 5$_{15}-4_{14}$ (2.3 $\times$ 10$^6$ cm$^{-3}$) \citep{shi15}. The remaining four observed H$_2$CO transitions have even higher energy levels and effective excitation densities. Indeed, the absorption signatures between 7.15 km s$^{-1}$ and 7.60 km s$^{-1}$ in HNC (4$-$3) are not evident in the observed H$_2$CO transitions, indicating that H$_2$CO probes material mainly in the close vicinity of 4A1. 


\begin{figure*}
\vspace{-1.4cm} 
\hspace{-.8cm} 
 \includegraphics[width=7.6in]{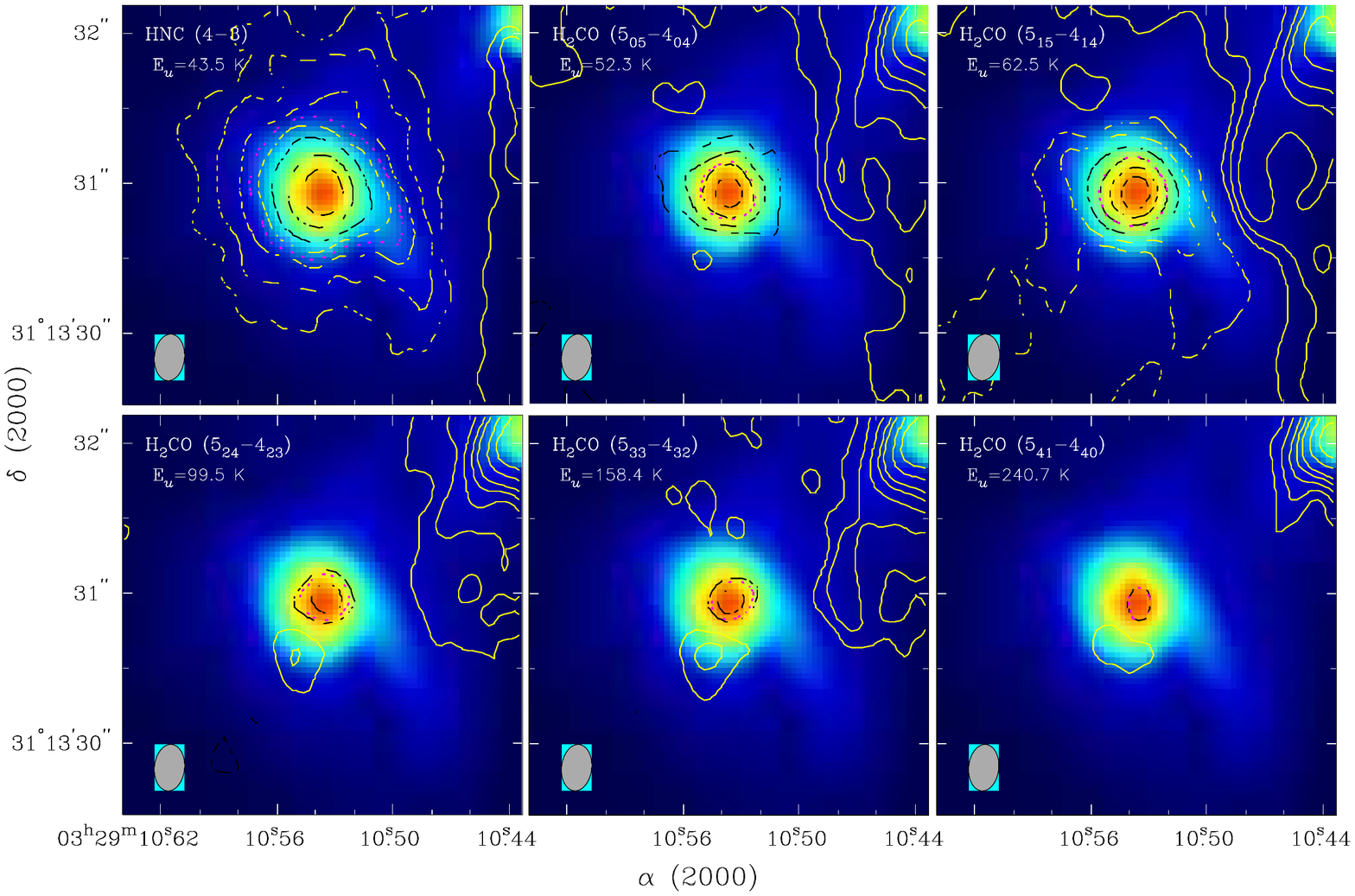}
 \vspace{-2.4cm} 
 \caption{Contour plots of the integrated intensity maps of HNC (4$-$3) and various H$_2$CO lines toward 4A1 integrated from 5.95 km s$^{-1}$ to 8.05 km s$^{-1}$. In each panel, the color scales present the 0.84~mm continuum shown in Figure \ref{cont} and the dark ellipse at the bottom left denotes the synthesized beam.  The transitions are labelled at the left top corner of each panel. Solid contours are at 3, 6, 10, 15, 20, 30, and 40 $\times$ 10 mJy beam$^{-1}$~km s$^{-1}$, and dashed contours indicate $-$3, $-$6, $-$10, $-$15, $-$20, $-$30 and $-$40 $\times$ 10 mJy beam$^{-1}$~km s$^{-1}$. In each panel, the red contour represents the 50\% of the peak absorption. \label{H2CO_absorption_img_4A1}} 
\end{figure*}

Figure~\ref{H2CO_absorption_img_4A1} presents the integrated intensity maps (integrated from 5.95 km s$^{-1}$ to 8.05 km s$^{-1}$) of HNC (4$-$3) and five observed H$_2$CO transitions toward 4A1. The absorption features of all five H$_2$CO transitions detected toward 4A1 are relatively compact, in particular for the high excitation lines. A weak emission component to the south-east of the centroid position is most likely related to the outflowing gas mentioned in the previous section. In contrast, the absorption seen in HNC (4$-$3) is fairly extended, as mentioned above. Among these H$_2$CO lines, there is no sign of a velocity gradient along the E-W direction, i.e., the axis perpendicular to the outflow associated with 4A1 shown in Figures \ref{outflow-chan} and \ref{outflow-int}. 

What are the physical properties of the H$_2$CO absorption gas? With multiple H$_2$CO transitions captured, we can estimate, to the zero-th order approximation, a ``mean/representative'' gas temperature using the rotation diagram analysis \citep{tur91} by assuming the observed H$_2$CO transitions in 4A1 are optically thin. We regard the continuum emission as the ``background'' in the radiative transfer equation, and the absorbing molecular line forms in a uniform ``foreground'' layer. Following the analysis presented in \citet{dip19}, a rotational temperature of 79 K and a H$_2$CO column density of 3$\times$10$^{14}$ cm$^{-2}$ toward the 4A1 centroid are estimated. For comparison, \citet{dip19} reported at the 4A1 center the rotational temperatures of $\sim$60 K and $\sim$140 K, respectively, for the observed CH$_3$OH and $^{13}$CH$_3$OH transitions, and the CH$_3$OH column density of $\sim$10$^{17}$ cm$^{-2}$. Assuming a typical fractional abundance of CH$_3$OH of 10$^{-6}$ \citep{her09}, the total column density of the absorbing complex organic molecules (COM) (atmosphere) layer is estimated to be at the level of 10$^{23}$ cm$^{-2}$. With a fractional abundance of H$_2$CO of 2$\times$10$^{-8}$ \citep{mar04}, our study suggests a column density of 3 $\times$10$^{22}$ cm$^{-2}$ for the absorbing layer, in agreement with that estimated from the CH$_3$OH transitions by \cite{dip19}.

While it is desirable to obtain a temperature map in the close vicinity of 4A1 by applying the rotation diagram analysis to a more extended area, the compact nature of the H$_2$CO absorption as well as the contamination from the outflowing gas make the applicable extent extremely limited. For example, the three high excitation H$_2$CO transitions (i.e., 5$_{33}-4_{32}$, 5$_{32}-4_{31}$, and 5$_{41}-4_{40}$) are not detected at a radius beyond 0\farcs25, equivalent to the synthesized beam size, away from the 4A1 center. At that radius, the detection of H$_2$CO 5$_{24}-4_{23}$ is also marginal, about 3$-$4 $\sigma$ only. Regarding the two low excitation H$_2$CO transitions (i.e., 5$_{05}$$-$4$_{04}$ and 5$_{15}$$-$4$_{14}$), although the absorption feature can be detected up to a radius of $\sim$0\farcs5, the contamination from outflowing gas (in particular along the N-S direction) makes the line profiles non-Gaussian. As a consequence, we simply extend the estimate of gas temperature to four representative locations 0\farcs25 to the N/S/E/W of 4A1.  With the three detected H$_2$CO transitions, the estimated gas temperatures are in range of 45 K to 60 K. The results appear to indicate a radial temperature gradient on the plane of sky.

\begin{figure*}
\vspace{-1.4cm} 
\hspace{-.8cm} 
 \includegraphics[width=7.6in]{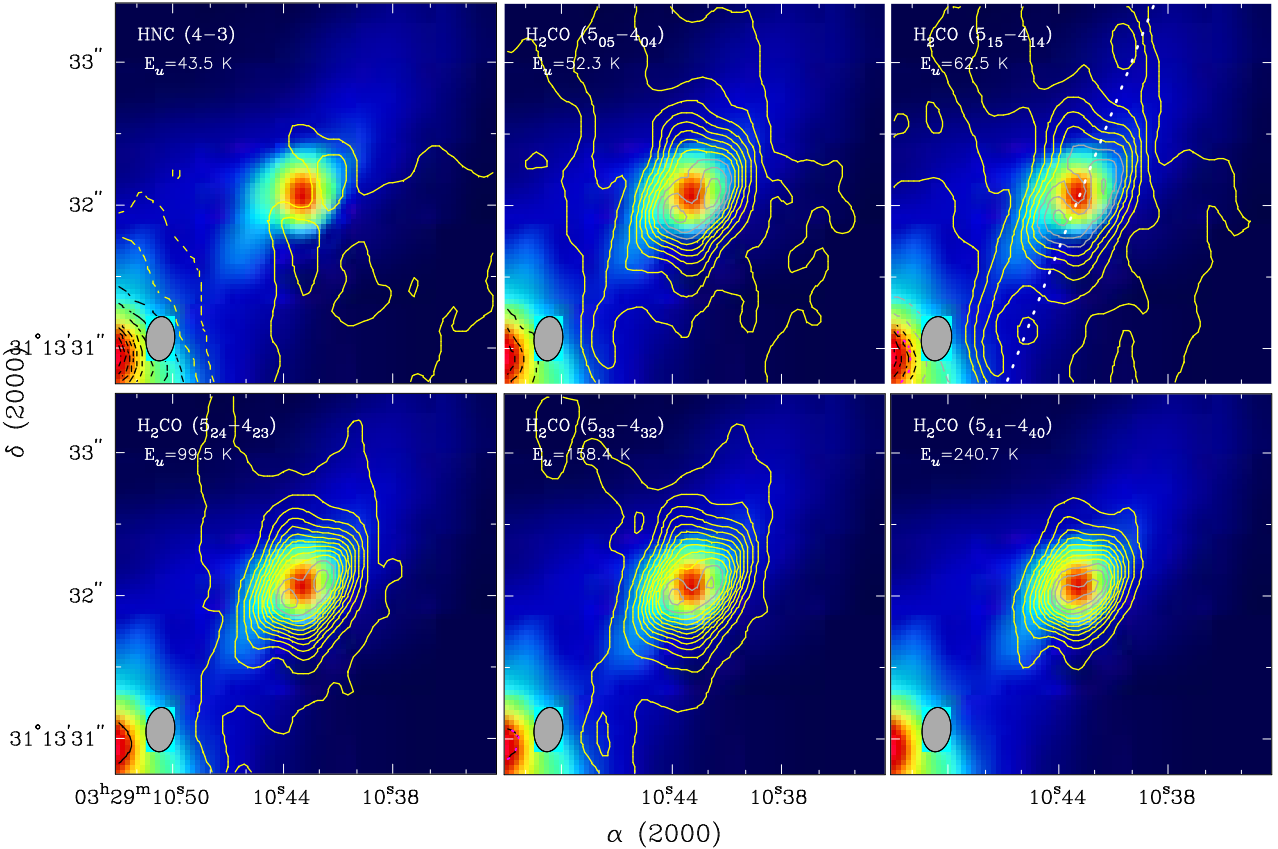}
 \vspace{-2.4cm} 
 \caption{Contour plots of the integrated intensity maps of HNC (4$-$3) and various H$_2$CO lines toward 4A2 integrated from 5.95 km s$^{-1}$ to 8.05 km s$^{-1}$. In each panel, the color scales present the 0.84~mm continuum shown in Figure \ref{cont} and the dark ellipse at the bottom left denotes the synthesized beam.  The transitions are labelled at the left top corner of each panel. Contours are at 4, 8, 12,... 48 $\times$ 10 mJy beam$^{-1}$~km s$^{-1}$.
 The white dotted  line shown in the panel of H$_2$CO 5$_{1,5}-4_{1,4}$ represents the axis of the position-velocity diagram plotted in Figure \ref{4A2-pvd}.
 \label{H2CO_absorption_img_4A2}} 
\end{figure*}

\begin{figure*}
\vspace{-.8cm} 
 \includegraphics[width=5.4in,angle=270]{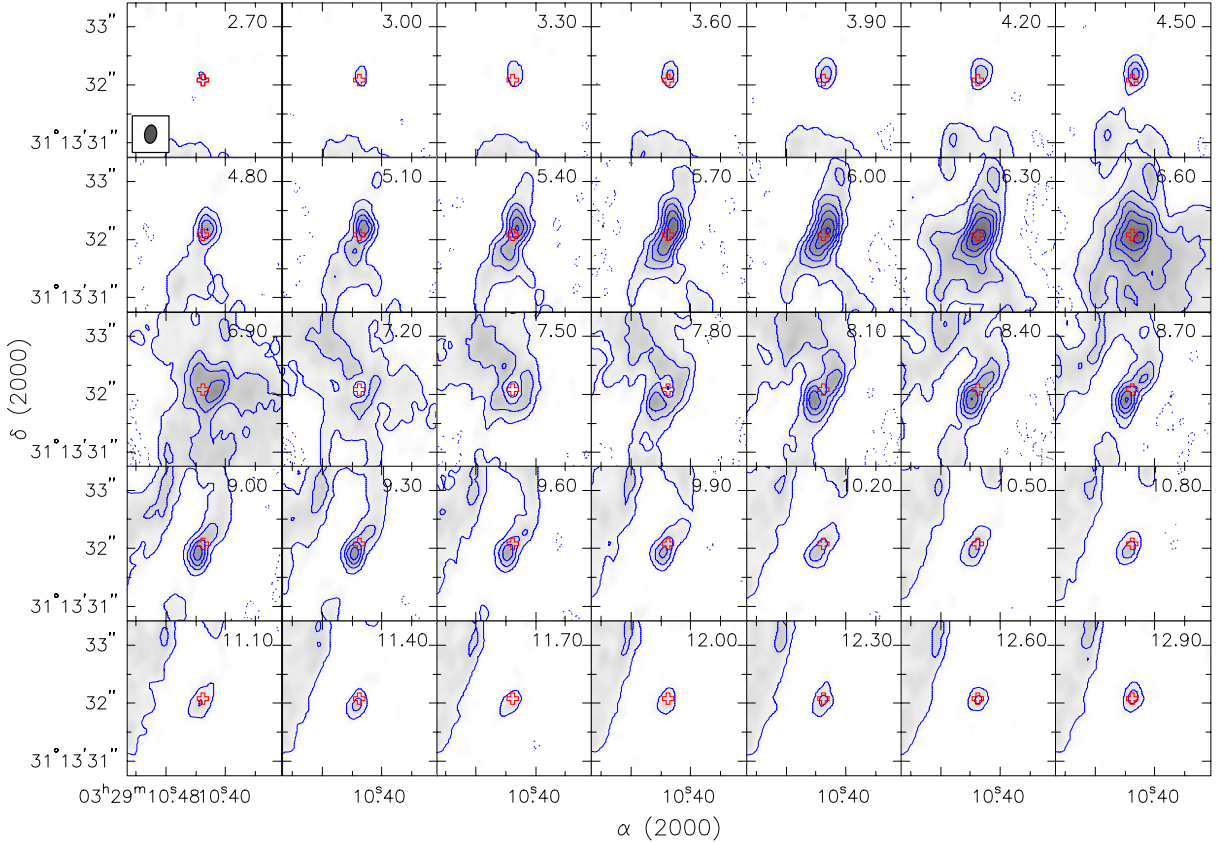}
 \vspace{-1.3cm} 
 \caption{Channel maps of H$_2$CO 5$_{15}-4_{14}$ zooming into the central region of 4A2.  The local standard of rest (LSR) velocity of each channel is indicated in the upper right corner of each panel. The LSR velocity of the 4A2 system is 7.00 km s$^{-1}$. Contours are at 4, 11, 18, 25, 32 and 39 $\times$ 7 mJy beam$^{-1}$. The dark ellipse at the bottom left of the first panel denotes the synthesized beam. In each panel, the red cross marks the position of the 4A2 compact component \citep{dip19}. At a scale of $\sim$1\arcsec, an outflow along the north-northwest to south-southeast direction can be discerned. \label{4A2-chan}}
\end{figure*}

\begin{figure}
\vspace{-.1cm} 
 \hspace{-2.2cm}
 \includegraphics[width=5.25in]{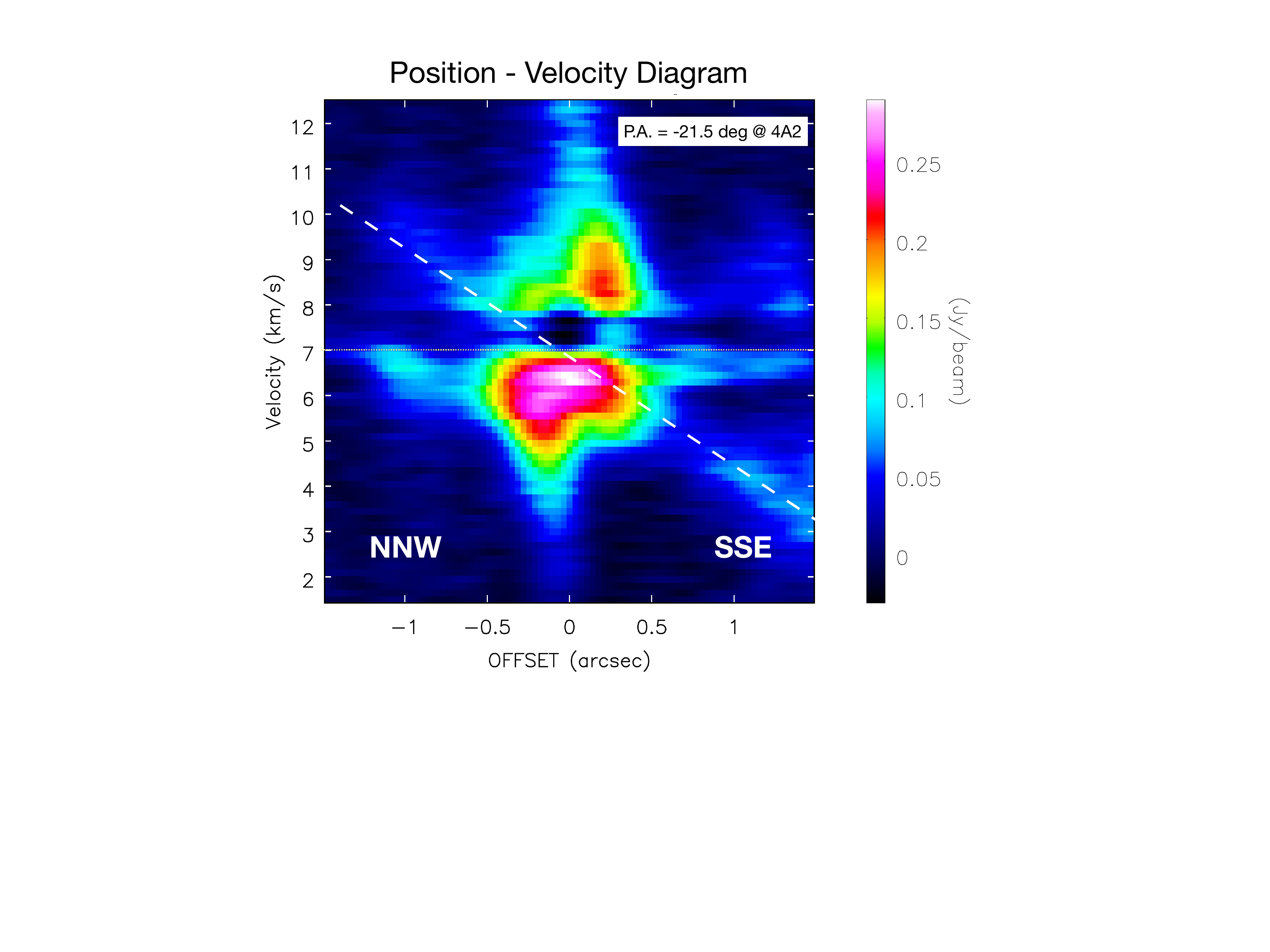}
 \vspace{-3.4cm} 
 \caption{Position-velocity diagram in H$_2$CO 5$_{1,5}-4_{1,4}$ along the P.A. of $-$21.5$^\circ$ centered at 4A2. The white dashed line labels the small scale outflow along the north-northwest to south-southeast direction. The dotted horizontal line denotes the 4A2 systemic velocity of 7.00 km s$^{-1}$.  \label{4A2-pvd}}
\end{figure}

\subsubsection{The Line Feature towards 4A2 \label{res_4a2}}
The \emph{right} panel of Figure~\ref{H2CO_spec_fitting} presents the observed spectra of the six H$_2$CO transitions and HNC (4$-$3) toward the 4A2 centroid.
Unlike a simple (absorption) Gaussian profile detected predominantly toward 4A1, the line profiles observed toward the 4A2 continuum centroid demonstrate more complicated features. The observed H$_2$CO as well as HNC (4$-$3) transitions all exhibit emission features, as shown in the \emph{right} panel of Figure \ref{H2CO_spec_fitting}. Absorption features are also detected toward the 4A2 center. Among these transitions, H$_2$CO 5$_{15}-4_{14}$ and HNC (4$-$3) demonstrate the inverse P-Cygni line profiles, i.e., absorption against continuum emission in redshifted velocity, and the remaining H$_2$CO lines exhibit asymmetric blue profiles. The observed line profile can be fit with a Gaussian shape if the low-velocity redshifted absorption channels are excluded. We again performed Gaussian profile fitting to extract the line parameters. The fitting profiles as well as the 4A2 systemic velocity of 7.00 km s$^{-1}$ estimated from the average fitted peak velocity are overlaid in the \emph{right} panel of Figure~\ref{H2CO_spec_fitting}. We note that the data points used in the fitting are marked as green points in Figure~\ref{H2CO_spec_fitting}. As summarized in Table~\ref{tb-line-fit}, the inferred linewidth is about 4.0 km s$^{-1}$, significantly broader than that estimated toward 4A1 of about 0.90 km s$^{-1}$ only. The estimated central line velocities are in agreement with each other, i.e., all in the range of 6.96 km s$^{-1}$ to 7.02 km s$^{-1}$. Another noticeable signature of the observed profiles is that all H$_2$CO transitions display similar peak intensities of $\sim$43 K (equivalent to $\sim$0.3 Jy beam$^{-1}$, or 85 K if the continuum emission included), indicating the optically thick nature of these H$_2$CO transitions. Such optically thick behavior toward 4A2 center has been revealed by \citet{dip19}, although the authors reported a slightly lower brightness temperature of 75 K (line + continuum), likely due to an intensity dilution in the spectral domain. We did not attempt the rotation diagram analysis to gauge the gas temperature of 4A2 due to the complex H$_2$CO emission/absorption profiles and its opaqueness at the emission peaks, which will lead to drastically overestimated gas temperature.

Figure~\ref{H2CO_absorption_img_4A2} presents the integrated intensity maps (integrated from 5.95 km s$^{-1}$ to 8.05 km s$^{-1}$) of HNC (4$-$3) and five observed H$_2$CO transitions toward 4A2. The integrated HNC (4$-$3) map does not reveal an evident feature as the emission and the foreground layer absorption roughly cancel out toward 4A2. The integrated intensity maps of H$_2$CO lines, on the other hand, all demonstrate strong emission features, with an elongated structure along a position angle of about $-$20$^{\circ}$. Since 4A2 is associated with a disk-outflow system \citep{cho10,san15}, the elongated structure can be naturally related to either the disk or the outflow. The direction of the elongated structure, however, is not aligned with that of the rotating disk around 4A2 which is oriented along a position angle of about 108.9$^\circ$ \citep{cho10}. The elongation is also slightly different from the observed outflow direction along the N-S direction shown in Figures~\ref{outflow-chan} and \ref{outflow-int}. Given the presence of the outflow precession in 4A2, however, the small-scale outflow in the close vicinity of 4A2 may not be necessarily aligned in the N-S direction. \\
To examine in detail the gas kinematics on small scales, we present in Figure~\ref{4A2-chan} the channel maps of H$_2$CO 5$_{1,5}-$4$_{1,4}$ zooming into the central (2\farcs5) region of 4A2 as well as in Figure \ref{4A2-pvd} the position-velocity diagram in H$_2$CO 5$_{1,5}-$4$_{1,4}$ also centered at 4A2. The most notable feature shown in Figure \ref{4A2-chan} is at a scale of $\sim$1\arcsec~an outflow oriented from north-northwest to south-southeast. The H$_2$CO gas kinematic structure, as revealed by the position-velocity diagram along the small-scale outflow direction (at a position angle of $-21^{\circ}.5$) shown in Figure \ref{4A2-pvd}, is however far more complicated. The position-velocity diagram exhibits a velocity gradient with red-shifted emission in the NNW and blue-shifted emission on the SSE side, which connect further to the large-scale N-S outflows. On the other hand, a separate emission component with a nearly reversed velocity gradient and which is even stronger in brightness is also revealed in Figure \ref{4A2-pvd}. While an outflow cone lying nearly in the plane of sky may result in both blue- and red-shifted gas emission along the same line-of-sight, the disparate brightness and extent of the observed emission features do not seem to support this picture. Most recently, Chuang et al. (submitted) witness similar morphological and kinematic structures in CO and SO emission $-$ a more extended emission with a velocity gradient aligned with the outflow and a compact ($\sim$1\arcsec~in diameter), perhaps flattened component with a reversed velocity gradient. They interpret the latter component as a rotating envelope and explain the bending outflow in the framework of magnetic fields misaligned from the initial rotation axis. Regardless of the interpretation, the elongated emission shown in Figure \ref{H2CO_absorption_img_4A2} is contributed by this compact component.

As mentioned above, the spectral profiles of H$_2$CO toward 4A2 exhibit either the inverse P-Cygni or asymmetric blue profiles, both hinting at the presence of inward motion in 4A2. The integrated intensity maps also show signatures of the infalling gas toward 4A2, as voids of H$_2$CO emission toward the 4A2 centroid can be clearly discerned in Figure~\ref{H2CO_absorption_img_4A2}. We note that such a void morphology can be seen in the integrated intensity map of low-excitation CH$_3$OH also (i.e., Figure 1\emph{b} in \cite{dip19}). The channel maps shown in Figure~\ref{4A2-chan} reveal an even clearer absorption feature from about 7.0 km s$^{-1}$ to 7.90 km s$^{-1}$, with a compact ($\sim$0\farcs5) ring-like emission structure symmetrically centered at 4A2. Such a ring-like morphology is detected not only in H$_2$CO 5$_{1,5}-$4$_{1,4}$ but also in other observed H$_2$CO transitions. With a systemic velocity of 7.0 km s$^{-1}$, the redshifted absorption signature most likely traces the inward motion of 4A2. We discuss in \S\ref{dis_4a2} the physical properties of the possible infalling gas in 4A2.

\begin{figure}
 \vspace{-.8cm} 
 \hspace{-1.2cm} 
 \includegraphics[width=4.2in]{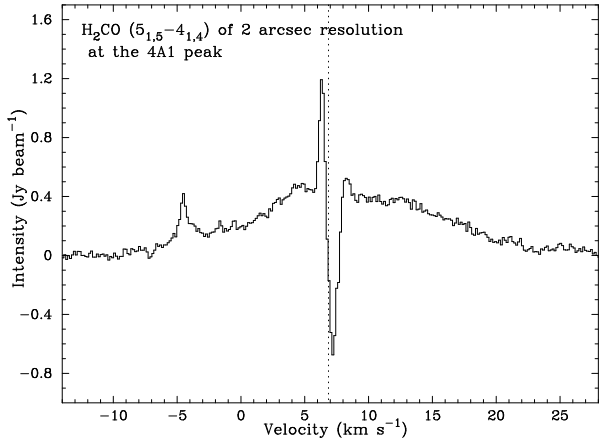}
 \vspace{-1.7cm} 
 \caption{Spectra of H$_2$CO 5$_{1,5}-4_{1,4}$ toward the central position of the 4A1 continuum with tapering the visibilities during imaging to 2\arcsec~resolution. The dotted line denotes the 4A1 systemic velocity of 6.86 km s$^{-1}$. Although the spectral profiles are the inverse P-Cygni like, the redshifted absorption feature at $\sim$7.2 km s$^{-1}$ and the blueshifted component peaking at $\sim$6.6 km s$^{-1}$ come from 4A1 and 4A2, respectively. \label{spec_taper}} 
\end{figure}

\section{Discussion \label{dis}}
\subsection{Gas Kinematics in 4A as a Whole \label{dis_4a1}}

The presence of the well-collimated bipolar outflows in 4A strongly implies that its embedded young binary stars are both still in an active accretion phase. Hence, one may expect to identify signatures of inward gas motion within the 4A region. Such signatures have been reported in previous studies. For example, one piece of evidence is the detection of the inverse P-Cygni profiles in H$_2$CO (3$_{12}-2_{11}$) with the PdBI observations at a resolution of $\sim$2\arcsec~\citep{dif01}. By matching the observed H$_2$CO profile with a two-layer radiative transfer model, \cite{dif01} deduced an infall velocity toward 4A of 0.68 km$^{-1}$.

To compare the H$_2$CO spectral features of our ALMA observations with the previous PdBI results, we downgrade the spatial resolution of our data cube to 2\arcsec, comparable to that of the PdBI observations, by tapering the visibilities during imaging. Indeed, we present in Figure \ref{spec_taper} an inverse P-Cygni signature in our ALMA 2\arcsec~H$_2$CO 5$_{15}-4_{14}$ spectral profile that is similar to the PdBI H$_2$CO 3$_{12}-2_{11}$ results. A broad line wing, most likely associated with the outflowing gas, in both red- and blue-shifted parts can be discerned in both transitions, though it is stronger and broader in our case. We note that the bright narrow blue-shifted emission peak at $\sim$ 6.6 km s$^{-1}$ comes from the contribution of 4A2.

Recently, Herschel WISH observations in multiple water transitions toward IRAS 4A also suggested infall motion in the IRAS 4A region as a whole. Through a series of studies \citep{jor02,mar04,kri12,mot13}, a detailed physical model of the region was constructed for interpreting the observations. The physical model parameters can be summarized as the following: density \emph{n} = 3.05 $\times$ 10$^9$ cm$^{-3}$ $\times$  (\emph{r}/33.5 au) $^{-1.8}$; temperature \emph{T} = 1.83  $\times$ (\emph{r}/100 au)$^{-2}$ + 47.5 $\times$ (\emph{r}/100 au)$^{-1}$ + 51.7 $\times$ (\emph{r}/100 au)$^{-0.4}$; infall velocity \emph{v}$_{inf}$ = \emph{v}$_{1000}$ $\times$ (\emph{r}/1000~au)$^{-0.5}$, where \emph{v}$_{1000}$ = 1.1 km s$^{-1}$; a turbulence velocity of 0.4 km s$^{-1}$; inner and outer boundary \emph{r}$_{in}$ = 33.2 au and \emph{r}$_{out}$ = 11200 au; a jump in formaldehyde fractional abundance with the jump occurring at the radius where the dust temperature reaches 100 K, i.e., \emph{X}(H$_2$CO) = 2 $\times$ 10$^{-10}$ for \emph{r} $>$ \emph{r}$_{100K}$, and 2 $\times$ 10$^{-8}$ for \emph{r} $<$ \emph{r}$_{100K}$, where \emph{r}$_{100K}$ $\sim$ 100 au with the above-mentioned temperature profile. In this particular model, the infall velocity at a radius of 1000 au (i.e., \emph{v}$_{inf}$ (\emph{r}) = \emph{v}$_{1000}$ $\times$ (\emph{r}/1000au)$^{-0.5}$) is estimated to be 1.1 km$^{-1}$ \citep{mot13}. We note that these modelling efforts, all based on single-dish observations which did not have sufficient angular resolution to resolve apart 4A1 and 4A2, only considered the IRAS 4A region as a single one-dimensional spherically symmetric collapsing core.

\begin{figure*}
\begin{center}
 \hspace{-1.5cm}
 \includegraphics[width=5.5in,angle=270]{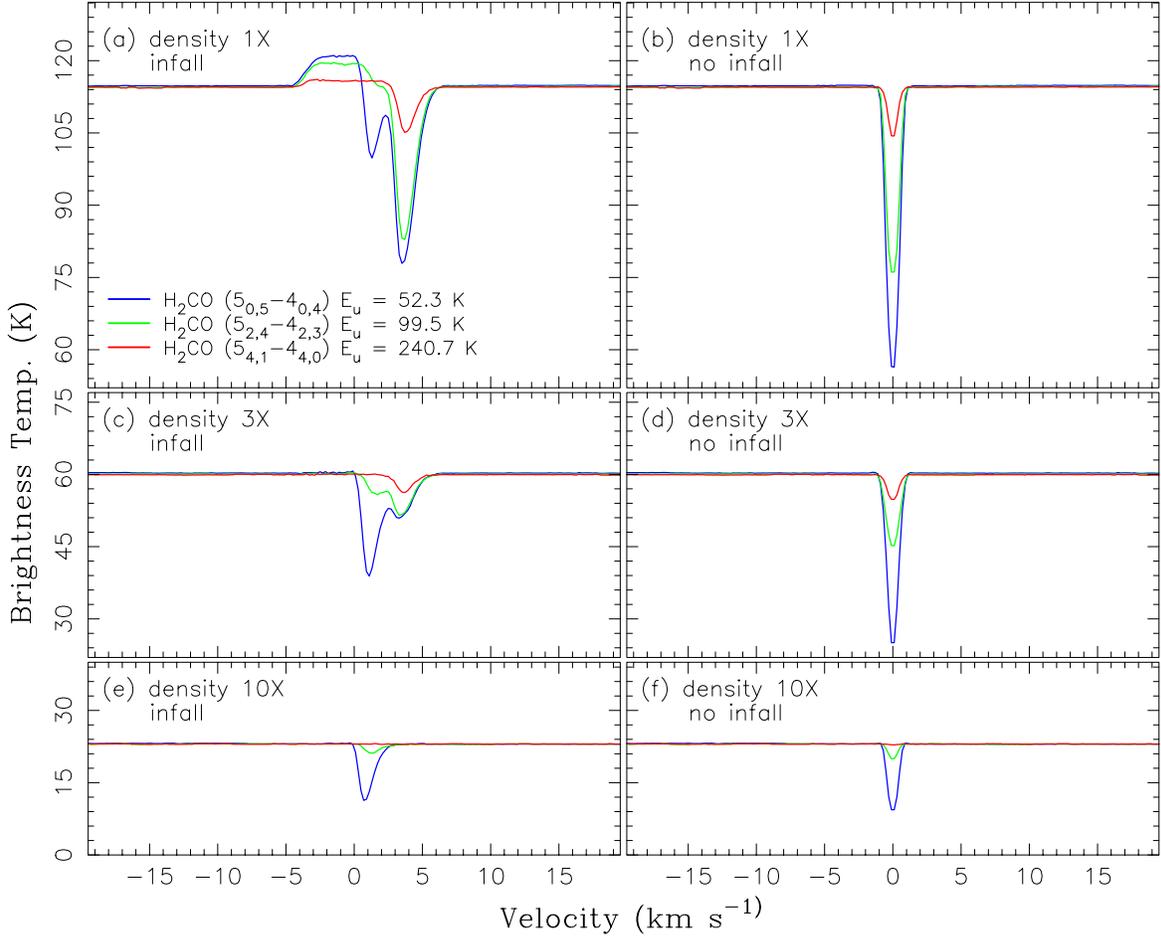}
 \vspace{-.7cm}
 \caption{Modeled spectra of H$_2$CO 5$_{05}$-4$_{04}$ (blue), 5$_{24}$-4$_{23}$ (green), and 5$_{41}$-4$_{40}$ (red) toward the central 0\farcs25~region of IRAS~4A with our SPARX calculations. Panel~\emph{a} presents simulated spectra of a toy 1D spherical collapsing model incorporating the physical parameters inferred from previous studies as alluded in \S \ref{dis_4a1}. Panels \emph{c} \& \emph{e}: same as panel \emph{a}, but increasing the original density profile to 3$\times$ and 10$\times$ values, respectively. Panels \emph{b}, \emph{d}, and \emph{f}: same as panels \emph{a}, \emph{c}, and \emph{e}, respectively, but no infalling-motion. The results demonstrate that high-excitation transitions are ideal tracers to probe faster infalling gas located closer to the central object. The absorption feature at 3$-$5 km s$^{-1}$ from faster inner infalling gas can be clearly discerned at 0\farcs2~resolution. \label{sparx-simulated-spectra}}
 \end{center}
\end{figure*}

\subsection{Radiative Transfer Modelling Exercises for 4A1}
When observed at a higher ($\sim$0\farcs25) angular resolution with ALMA, the H$_2$CO and HNC transitions toward the 4A1 center, as mentioned in \S \ref{res_4a1}, display distinctly different behavior from those acquired by single dish observations. They are mainly Gaussian-like profiles predominantly in absorption, with typical linewidths of only $\sim$1 km s$^{-1}$. Furthermore, contrary to the expectation from the physical model outlined above, based on which a high velocity (3$-$5 km s$^{-1}$) infalling signature should be present, the observed profiles show no sign of (redshifted) infalling gas with velocity $\gtrsim$ 0.5 km s$^{-1}$.

What is the cause of this lack of high-velocity gas infall signature in H$_2$CO toward 4A1? As illustrated in \S \ref{resultcont}, the dust continuum at 0.84~mm toward the 4A1 centroid is optically thick. Is the opaque dust emission preventing the detection of high-velocity infalling gas which is expected to be located at the inner region of the 4A1 envelope?
To investigate these effects further, we resort to radiative transfer (RT) modelling exercises.

We perform non-LTE radiative transfer calculations with the SPARX\footnote{SPARX, which stands for Simulation Package for Astrophysical Radiative Transfer, is a program for radiative transfer calculation of molecular lines and continuum using a parallelized accelerated Monte Carlo algorithm over full 3-D nested grids. \citep{liu19}} package over the multiple H$_2$CO transitions. As an initial trial step, we adopt a toy 1D spherical collapsing core model of IRAS 4A incorporating the physical parameters inferred from previous studies as alluded above. 
To take the possible spatial filtering resulted from the present ALMA observations into account, we simulated ALMA observations at its C43-3 and C43-5 configurations, same as those used in our ALMA observations by using the SPARX simulated datacubes as input model images. The beam size of the mock ALMA observations is almost the same as that of our observed ALMA results. Panel \emph{a} of Figure~\ref{sparx-simulated-spectra} shows the simulated spectra of the three (para-)H$_2$CO transitions (i.e., 5$_{05}-4_{04}$, 5$_{24}-4_{23}$, and 5$_{41}-4_{40}$) toward the continuum peak in the datacube obtained from the mock ALMA observations.
The spectra of these three transitions exhibit both blue-shifted emission features primarily associated with the infalling gas on the far side and red-shifted absorption features associated with the infalling gas on the near side.
For the low-excitation one (i.e., 5$_{05}-4_{04}$), the simulated absorption signature is kinematically distinct in nature, with the ``low-velocity'' (\emph{v}~$\sim$ 1$-$3 km s$^{-1}$) feature mainly rising from gas located at the outer/cool (T $<$ 100 K) envelope at a relatively small infallng speed and the ``high-velocity'' (\emph{v}~$\sim$ 3$-$5 km s$^{-1}$) feature originating from gas in the inner/warm (T $>$ 100 K) collapsing core where H$_2$CO fractional abundance is enhanced. 
The simulated spectra of the two high-excitation transitions, on the other hand, display absorption signatures in ``high-velocity'' only, due to insufficient pumping of H$_2$CO molecules up to high-excitation states in the outer/cool envelope.

The modeled spectral profiles, however, are significantly different from our observed ALMA results. Additionally, the modeled continuum brightness temperature is about a factor of 2 higher than the observed value. For a better understanding of how the adopted parameters affect the simulated spectral features, we performed a further set of RT calculations by varying the input density and infalling velocity. 
With the original adopted density parameter, the modeled dust continuum emission has a brightness temperature T $\sim$ 110 K, higher than the observed value of 60 K. We tried two other high density cases by increasing the original density profile to 3$\times$ and 10$\times$ values. These simulated spectra, as a consequence, reflect the effects of dust opacity variations in both continuum brightness and spectral features. The modeled continuum brightness shows a decreasing trend with increasing density (and hence column density). That is, the peak continuum brightness \emph{T}$_{b}$ is about 110 K, 60 K, and 23 K, respectively, for the calculations with 1$\times$,  3$\times$, and 10$\times$ in density. The result is consistent with the fact that an increasing dust/gas density leads to the dust continuum optically thick surfaces shifting to the outer/cooler regions.
Regarding the spectral features, as shown in Figure \ref{sparx-simulated-spectra}, the contrast between high- (\emph{v}~$\sim$ 3$-$5 km s$^{-1}$) and low- (\emph{v}~$\sim$ 1$-$3 km s$^{-1}$) velocity absorption in these simulated spectra also demonstrates a trend with input density. The ``high-velocity'' feature is the dominant component in the 1$\times$ density case, and vanishes in the 10$\times$ density case. In contrast, the ``low-velocity'' feature is the minor component in the 1$\times$ density case, but becomes the major one in the 3$\times$ density case as the faster infalling material closer to the central star becomes optically thicker and thus contributes less to the absorption. Since ``high-velocity'' absorption features trace infalling gas located at the inner/warm envelope, the decreasing trend of the ``high-velocity'' absorption depth with dust/gas density is the result of increasing dust obscuration of the inner/warm envelope. Meanwhile, no clear dust opacity effect can be discerned on the modeled ``low-velocity'' absorption signatures. Finally, we also set up RT models with non-infalling motion in the 4A1 envelope for comparison purposes. The modeled spectral profiles in these cases (shown in panels \emph{b}, \emph{d}, and \emph{f} of Figure \ref{sparx-simulated-spectra}) display absorption features only at the systemic velocity, as expected.

In short, our RT calculations indicate that although dust opacity plays an essential role in obstructing the detection of ``high-velocity'' infalling gas, ``low-velocity'' absorption feature should be readily captured when the infall profile following the fiducial model is considered. Our observed H$_2$CO profiles, however, lack such signatures of both high- and low-velocity infalling motions, which is puzzling. We discuss possible resolutions to this interesting puzzle next. 

\subsection{The physical structures of 4A1 \label{4A1_structure}}
The discrepancies between the observed and the modelled spectra of 4A1 suggest that the physical parameters inferred from earlier low-resolution observations and used in the toy models cannot be directly extrapolated down to the inner region of 4A1. We discuss below possible reasons for such differences. Firstly, the actual inward motion may be significantly slower or the velocity profile is much shallower than that inferred by \cite{mot13}. The simulated spectra of the 3$\times$ density case, which reproduces well the observed continuum brightness, exhibits ``high-velocity'' absorption features up to 4.0 km s$^{-1}$ in relative velocity, arising from gas located at 75 au. Since our observations show no sign of (redshifted) infalling gas with velocity $\gtrsim$ 0.5 km s$^{-1}$, we estimate the radial inward motion at a radius of 75 au should be at least a factor of 8 slower than that extrapolated from the velocity structure deduced by \cite{mot13}. 
Using the parameters described above at the layer of 75 au, we further constrain the mass infall rate in 4A1 to be (at most) around 3$\times$10$^{-5}$ \emph{M}$_{\odot}$ year$^{-1}$, which is about a factor of 5 smaller than that obtained by \cite{mot13} on the 1000 au scale.

In addition, the one-dimensional collapsing core model approximation most likely has to break down. On the large scale, there are multiple energetic bipolar molecular outflows. In the inner envelope, the existence of a rotating disk in the close vicinity of 4A1 is expected for launching the bipolar outflows. The presence of outflows and disks would call for a 2/3-D description of the core/envelope structure surrounding 4A1. In this case, the actual gas accretion flowing onto the disk may not lie along the line-of-sight and hence escape detection through line absorption against the continuum.

What is the origin of the H$_2$CO absorption material seen towards 4A1, then?
\citet{dip19} proposed two possible scenarios to account for their observational absorption signatures of COMs, including CH$_3$OH, CH$_3$OD, and CH$_3$CHO, toward 4A1. 
In their Scenario I, the COM absorption lines arise from a hot-corino-like atmosphere at the surface of an optically thick circumstellar disk around 4A1 viewed closely face on.
Under this scenario, a high accretion rate in the disk is actually required. 
The active accretion helps to produce a hot and bright continuum background in the inner disk required by the absorption feature.
With comparable temperature and column density estimated from the rotational diagram analysis described in \S \ref{res_4a1}, it is possible that the absorptions of CH$_3$OH and H$_2$CO both arise from a hot-corino-like atmosphere. The nearly face-on disk configuration in this picture naturally breaks down the 1D assumption, and the accretion flow onto the disk may be channeled mostly close to the plane of sky as suggested above.

In Sahu et al.'s Scenario II, the absorption features arise from different layers of a temperature-stratified, dense envelope. 
All or part of the absorption features toward 4A1 may be optically thick and hence their brightness at the absorption line core is equal to the foreground absorbing gas temperature.
The absorption lines with different excitation temperatures form at different layers in a temperature-stratified envelope --- the lower excitation line saturates at the outer cooler layer, while the higher excitation transition (not excited at the outer layer) becomes optically thick in the relatively inner and hotter region. The 1D RT calculations with no infall motion discussed above resemble the physical nature of this Scenario II. We note here that the rotational diagram analysis described in \S \ref{res_4a1} requires the assumption of all observed transitions arising from the same region and hence it contradicts Scenario II. High-resolution observations of disk tracers at low frequency may help to discriminate the two scenarios if the disk is not totally face-on but slightly inclined.

\subsection{Gas Kinematics in 4A2 \label{dis_4a2}}
As illustrated in \S \ref{res_4a2}, the line profiles for the H$_2$CO transitions toward the 4A2 continuum centroid are more complicated. They exhibit both emission and absorption with characteristics resembling inverse P-Cygni or asymmetric blue-skewed profiles. Based on the position-velocity diagram (Figure \ref{4A2-pvd}), the emission part of the spectra (with a linewidth of $\sim$ 4~km s$^{-1}$) originates from a compact ($\sim$1\arcsec), perhaps flattened envelope. We fit the profile of this H$_2$CO emission component, as described in \S \ref{res_4a2}, and assigned the average fitted peak velocity of 7.00 km s$^{-1}$ as the systemic velocity of 4A2. This systemic velocity is consistent with the results of previous studies \citep{lop17,dip19}. In particular, the line profiles of HCOOCH$_3$, CH$_3$OCH$_3$, HCOOCH, HNCO etc. toward 4A2 reported by \cite{lop17} at 0\farcs$5-$1\arcsec~resolution are mostly Gaussian-like, with fitted peak velocities predominantly in range of 6.8 km s$^{-1}$ to 7.1 km s$^{-1}$.

Meanwhile, there is also an intriguing feature in the H$_2$CO integrated intensity maps toward 4A2.
The position angle of the integrated emission appears to rotate from about $-$21.5$^{\circ}$ for the outer ($\sim$1\arcsec~scale) lower level contours to about $-$60$^{\circ}$ (or 120$^{\circ}$) for the inner ($\sim$0\farcs4~scale) higher level contours. Coincidentally, a rotating disk associated with 4A2 at a position angle of $\sim$110$^\circ$ was suggested by the VLA subarcsec (0\farcs3) observations in NH$_3$ (2,2) and (3,3) \citep{cho10}. Assuming a Keplerian rotation, those authors estimated a rotation velocity of 1.8 km s$^{-1}$ at a radius of 25 au (after scaling their adopted distance to from 235 au to 293 au), which implies a central object of about 0.08 \emph{M}$_\sun$.

\begin{figure}
 \vspace{-.8cm} 
 \hspace{-1.5cm} 
 \includegraphics[width=4.3in]{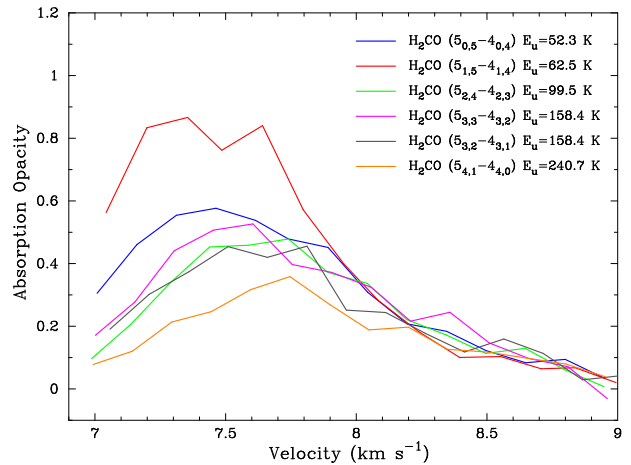}
 \vspace{-1.8cm} 
 \caption{Absorption opacities of the six observed H$_2$CO transitions as a function of gas velocities (\emph{V}$_{\rm LSR}$) toward the 0.84~mm continuum peak of the 4A2. Given a systemic velocity of 7.0 km s$^{-1}$, there appears to be a trend that the relative velocity of the peak opacity increases from 0.4 km s$^{-1}$ to 0.8 km s$^{-1}$ with the line excitation. The trend is consistent with the picture that the high-excitation line traces better the higher infall velocities closer to the central young star. \label{4a2_tau}} \end{figure}

\begin{figure*}
\vspace{-1.3cm}
\hspace{1.cm}
\includegraphics[width=6.5in]{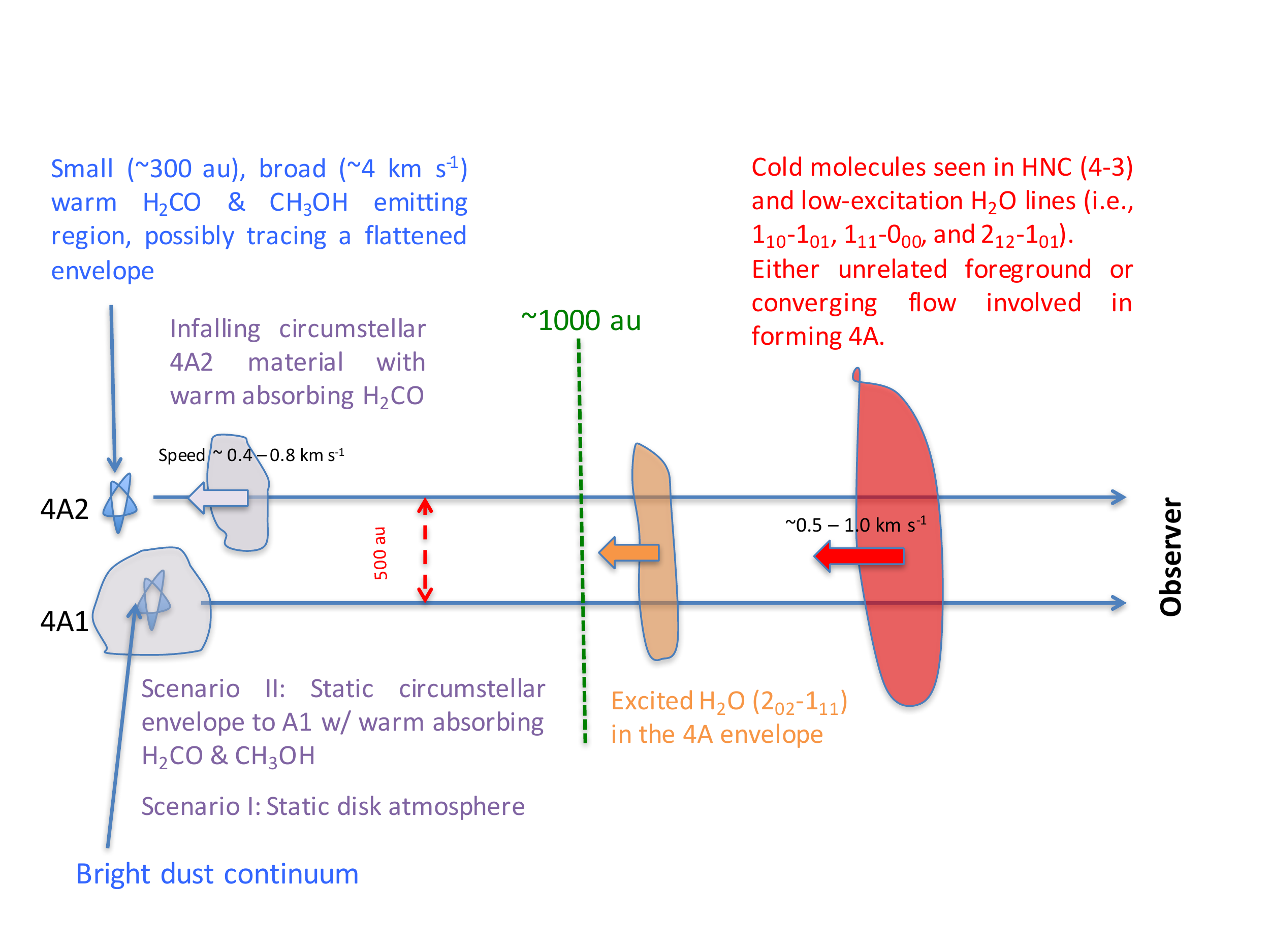}
\vspace{-.8cm} 
 \caption{Schematic illustration of the IRAS 4A system inferred from the present ALMA observations as well as the results presented by \cite{mot13} and \cite{dip19}. \label{IRAS4A1A2_Cartoon}} \end{figure*}

Similar to the expectation for 4A1, one may also anticipate evidence for inward gas motion toward 4A2, given its association with a disk/outflow system. On one hand, with this adopted systemic velocity, the absorbing dips in the H$_2$CO spectra are indeed skewed toward the red-shifted side and likely signatures of inward motion. On the other hand, these absorbing dips are relatively narrow, extending up to around \emph{V}$_{\rm LSR}$ $\sim$ 8 km~s$^{-1}$. As for the case for 4A1, these speeds are significantly slower than the anticipated inflow speed should the 1D collapsing core model be extrapolated down to this scale. The lack of a proper description for the complex mixture of disk, outflow, and envelope components around 4A2 prevents us from formulating a full RT modeling experiment.
Here we try to interpret the observed absorption/skewed profiles with another approach. We assume the fitted Gaussian profiles shown in the \emph{right} panel of Figure \ref{H2CO_spec_fitting} represent the spectral feature of a ``model" without absorption. As a consequence, we can estimate the absorption opacity using the intensity difference between the model and the observed line profiles. With similar peak intensities of $\sim$85 K, most H$_2$CO transitions are likely optically thick at the center velocity. The absorption opacity, however, could be significantly smaller, in particular, off the line center. We estimate the line opacity of the absorption layer as a function of velocity for all the observed H$_2$CO lines by assuming the optically thin case: 
$\tau$(\emph{v}) $\simeq$~(1 $-$ exp($-$ $\tau$(\emph{v}))) = ((\emph{I$_{line-model}$}(\emph{v})+\emph{I$_{cont}$})$-$(\emph{I$_{line-obs}$}(\emph{v})+\emph{I$_{cont}$}))/(\emph{I$_{line-model}$}(\emph{v})+\emph{I$_{cont}$}), where \emph{I$_{cont}$}, \emph{I$_{line-obs}$}, and \emph{I$_{line-model}$} are the observed continuum intensity, the observed line intensity, and the intensity of the fitted Gaussian spectra shown in Figure \ref{H2CO_spec_fitting}, respectively. Figure \ref{4a2_tau} presents the estimated absorption opacities of the six observed H$_2$CO transitions as a function of gas velocities (\emph{V}$_{\rm LSR}$) toward the 0.84~mm continuum peak of the 4A2. As shown in Figure \ref{4a2_tau}, for all the transitions the largest opacities are in the range of 7.4 km s$^{-1}$ to 7.8 km s$^{-1}$. With the assigned systemic velocity of 7.0 km s$^{-1}$, interestingly, there appears to be a trend that the relative velocity of the peak opacity increases from 0.4 km s$^{-1}$ to 0.8 km s$^{-1}$ with the line excitation. The trend is consistent with the picture that the high-excitation line traces better the higher infall velocities closer to the central young star. 


Given that the excitations of the observed H$_2$CO transitions are in range of 50 K $-$ 240 K, we adopt a gas temperature of $\sim$100 K for the observed infalling material in 4A2. If the density and temperature structures of the 4A2 inner envelope are similar to those inferred from earlier low-resolution observations and used in our toy models mentioned in \S\ref{dis_4a1}, then the observed inward material of \emph{T}$\sim$100 K is located at a layer of 100 au with a density of 4.3$\times$10$^{8}$ cm$^{-3}$. Together with an infall velocity of 0.4 $-$ 0.8 km s$^{-1}$, we anticipate a mass infall rate of 3.1$-$6.2 $\times$ 10$^{-5}$ \emph{M}$_\sun$ year$^{-1}$ at the layer of 100 au in 4A2. For comparison, \cite{mot13} estimated a mass infall rate of 1.54 $\times$ 10$^{-4}$ \emph{M}$_\sun$ year$^{-1}$ for the 4A system on the 1000 au scale.

\subsection{The Configuration of the IRAS 4A System \label{sec:dis-configuration}}
Figure \ref{IRAS4A1A2_Cartoon} presents a schematic illustration of the IRAS 4A system. The cartoon provides connections between the observational results and the interpretations presented in this paper and in the literature, especially \cite{mot13} and \cite{dip19}. On the large scale, the available data indicate that there are (at least) two types of absorbers. The first one is a cold, low-excitation layer of molecular gas located between the IRAS 4A system and the observer, with an extent covering both 4A1 and 4A2. This layer can be discerned in our ALMA HNC (4$-$3) results and the low-excitation H$_2$O transitions (i.e., 1$_{10}-1_{01}$, 1$_{11}-0_{00}$, and 2$_{12}-1_{01}$) reported by \cite{mot13}. It appears that this component is moving towards the IRAS 4A system at a speed of order $\sim$0.5$-$0.8 km s$^{-1}$, i.e., supersonically with a Mach number of about 3$-$4 for a gas temperature of $\sim$10 K. This layer could be part of a converging flow that is involved in the formation of the dense gas that collapsed into the IRAS 4A system or just an independent gas layer. The converging flow may result from the collision between the NGC1333 main cloud and a turbulent cell, as introduced by \citet[][see their Fig.18]{Dha19}. The second absorber is identified in H$_2$O (2$_{02}-1_{11}$), as part of the envelope of the IRAS 4A system \citep{mot13}. The authors conclude that this water line comes from a distance of $\gtrsim$1000~au from the 4A system, suggesting that this water layer should surround both 4A1 and 4A2 given their (projected) separation of $\sim$500 au only. 

On the small scale, simple, symmetric spectral profiles in absorption toward the 4A1 dust continuum centroid have been detected in multiple H$_2$CO transitions presented here as well as CH$_3$OH and several COMs reported by \cite{dip19}. With no sign of (redshifted) infalling gas with velocity $\gtrsim$ 0.5 km s$^{-1}$, we estimate the radial inward motion at a radius of 75 au should be at least a factor of 8 slower than that extrapolated from the velocity structure deduced by \cite{mot13}. Two possible scenarios have been proposed by \cite{dip19} to account for their observational absorption signatures of COMs toward 4A1. In their Scenario I, the COM absorption lines arise from a hot-corino-like atmosphere at the surface of an optically thick circumstellar disk around 4A1 viewed relatively face-on. For the Scenario II, the absorption features arise from different layers of a temperature-stratified, dense envelope. 

4A2 is a known hot-corino like source \citep[][and reference therein]{dip19}. Given similar peak intensities, most observed transitions are likely optically thick at the center velocity under 0\farcs25 resolution \citep[][and this study]{dip19}. With the adopted systemic velocity of 7.0 km s$^{-1}$, the absorbing dips in the H$_2$CO spectra are skewed toward the red-shifted side and hence are likely signatures of inward motion. Similar to the case of 4A1, the observed infall velocity of 0.4 km s$^{-1}$ $-$ 0.8 km s$^{-1}$ is significantly slower that extrapolated from the velocity structure deduced by \cite{mot13}. We estimate a mass infall rate of 3.1$-$6.2 $\times$ 10$^{-5}$ \emph{M}$_\sun$ year$^{-1}$ at the layer of 100 au in 4A2. This rate is somewhat less than, but comparable to that inferred by \cite{mot13} on the 1000 au scale from H$_2$O absorption. The presence of infall in the inner envelope of the 4A system is consistent with the hour-glass shaped magnetic field observed by \cite{gir06}, since the infalling gas is expected to drag the magnetic field lines into a highly pinched configuration \citep[e.g.,][]{gal93}. 

\section{Summary}
We report ALMA observations of NGC 1333 IRAS 4A, a young low-mass protostellar binary, in multiple H$_2$CO transitions as well as HNC (4$-$3) at a resolution of 0$\farcs$25 ($\sim$70 au) to investigate the gas kinematics of 4A1 and 4A2. Our main results can be summarized as follows.

1. On the large angular scale ($\sim$10$\arcsec$), 4A1 and 4A2 each display a well-collimated outflow along the N-S direction, and an S-shaped morphology can be discerned in the outflow powered by 4A2. 

2. With deep absorption between 7.15 km s$^{-1}$ and 7.60 km s$^{-1}$ covering the whole extent of the 0.84 mm continuum, our HNC (4$-$3) results confirm the existence of a previously inferred redshifted foreground layer. 

3. On the small angular scale ($\sim$0\farcs3), 4A1 and 4A2 exhibit distinct spectral features toward the continuum centroid, with 4A1 showing simple symmetric profiles predominantly in absorption and 4A2 demonstrating rather complicated profiles in emission as well as in absorption. 

4. With Gaussian-like profiles and typical line widths of 1 km s$^{-1}$, the observed H$_2$CO profiles toward 4A1 show no sign of (redshifted) infalling gas with velocity $\gtrsim$0.5 km s$^{-1}$. We perform non-LTE radiative transfer calculations with the SPARX package incorporating the physical parameters inferred from previous studies. The modelled profiles indicate that although dust opacity toward 4A1 plays an essential role in obstructing the detection of ``high-velocity'' (3$-$5 km s$^{-1}$) infalling gas, a ``low-velocity'' (1$-$2 km s$^{-1}$) absorption feature should be readily detected. The discrepancies between the observed and the modelled spectra of 4A1 suggest that the physical parameters inferred from earlier low-resolution observations and used in the toy models cannot be directly extrapolated down to the inner region of 4A1. We constrain the mass infall rate in 4A1 to be (at most) around 3$\times$10$^{-5}$ \emph{M}$_{\odot}$ year$^{-1}$, which is about a factor of 5 smaller than that obtained by \cite{mot13} on the 1000 au scale.

5. For the kinematics of the 4A2 inner envelope, the absorbing dips in the H$_2$CO spectra are indeed skewed toward the redshifted side and likely signatures of inward motion. These absorbing dips are relatively narrow, extending up to around \emph{V}$_{\rm LSR}$ $\sim$ 8 km~s$^{-1}$. As in the case for 4A1, these speeds are significantly slower than the anticipated inflow speed should the toy model be extrapolated down to this scale. We estimate a mass infall rate of 3.1$-$6.2 $\times$ 10$^{-5}$ \emph{M}$_\sun$ year$^{-1}$ at the layer of 100 au in 4A2. For comparison, \cite{mot13} estimated a mass infall rate of 1.54 $\times$ 10$^{-4}$ \emph{M}$_\sun$ year$^{-1}$ for the 4A system on the 1000 au scale. 

\acknowledgments  This paper makes use of the following ALMA data: ADS/JAO.ALMA$\#$2015.1.00147.S. ALMA is a partnership of ESO (representing its member states), NSF (USA) and NINS (Japan), together with NRC (Canada), MOST and ASIAA (Taiwan), and KASI (Republic of Korea), in cooperation with the Republic of Chile. The Joint ALMA Observatory is operated by ESO, AUI/NRAO and NAOJ. We thank the anonymous referee for the constructive suggestions. YNS thanks T.-H. Hsieh for providing the code used to infer spectral parameters and to make Figure 4. YNS acknowledges the support by the Minister of Science and Technology of Taiwan (MOST 107-2119-M-001-041 and MOST 108-2112-M-001-048). ZYL is supported in part by NASA 80NSSC18K1095 and NSF AST-1716259, 1815784, and 1910106. S.T. acknowledges a grant from JSPS KAKENHI Grant Number JP18K03703 in support of this work. This work was supported by NAOJ ALMA Scientific Research grant No. 2017-04A.


\begin{thebibliography}{}

\bibitem[Belloche et al.(2006)]{bel06} Belloche, A., Parise, B., van der Tak, F. F. S., et al. 2006, \aap, 454, L51
\bibitem[Blake et al.(1995)]{bla95} Blake, G. A., Sandell, G., van Dishoeck, E. F., et al. 1995, \apj, 441, 689
\bibitem[Bottinelli et al.(2004)]{bot04} Bottinelli, S., Ceccarelli, C., Lefloch, B., et al. 2004, \apj, 615, 354
\bibitem[Chen et al.(2013)]{che13} Chen, X., Arce, H. G., Zhang, Q., et al. 2013, \apj, 768, 110
\bibitem[Choi et al.(2004)]{cho04} Choi, M., Kamazaki, T., Tatematsu, K., \& Panis, J.-F. 2004, \apj, 617, 1157 
\bibitem[Choi(2005)]{cho05} Choi, M. 2005, \apj, 630, 976 
\bibitem[Choi, Tatematsu \& Kang(2010)]{cho10} Choi, M., Tatematsu, K. \& Kang, M. 2010, \apjl, 723, L34
\bibitem[Choi et al.(2011)]{cho11} Choi, M., Kang, M., Tatematsu, K., Lee, J.-E., \& Park, G. 2011, \pasj, 63, 1281
\bibitem[Cox et al.(2015)]{cox15} Cox E. G., Harris R. J., Looney L. W., et al. 2015 \apjl, 814, L28
\bibitem[Dhabal et al.(2019)]{Dha19} Dhabal, A., Mundy, L.~G., Chen, C.-. yu ., et al.\ 2019, \apj, 876, 108
\bibitem[Di Francesco et al.(2001)]{dif01} Di Francesco, J., Myers, P. C., Wilner, D. J., Ohashi, N., \& Mardones, D. 2001, \apj, 562, 770
\bibitem[Evans et al.(2015)]{eva15} Evans N. J. II, Di Francesco J., Lee J.-E., et al. 2015, \apj, 814, 22
\bibitem[Galli, \& Shu(1993)]{gal93} Galli, D., \& Shu, F.~H.\ 1993, \apj, 417, 243
\bibitem[Girart, Rao, \& Marrone(2006)]{gir06} Girart, J. M., Rao, R., \& Marrone, D. P. 2006, Science, 313, 812
\bibitem[Gon\c{c}alves, Galli, \& Girart(2008)]{gon08} Gon\c{c}alves, J., Galli, D., \& Girart, J. M. 2008, \aap, 490, L39
\bibitem[Gueth \& Guilloteau(1999)]{gue99} Gueth, F., \& Guilloteau, S. 1999, \aap, 343, 571
\bibitem[Herbst \& van Dishoeck (2009)]{her09} Herbst, E. \& van Dishoeck, E.F., 2009, \araa, 47, 427 
\bibitem[J$\o$rgensen et al.(2002)]{jor02} J$\o$rgensen, J. K., Sch\"{o}ier, F. L., \& van Dishoeck, E. F. 2002, \aap, 389, 909
\bibitem[J$\o$rgensen et al.(2007)]{jor07} J$\o$rgensen, J. K., Bourke, T. L., Myers, P. C., et al. 2007, \apj, 659, 479
\bibitem[Kristensen et al.(2012)]{kri12} Kristensen, L. E., van Dishoeck, E. F., Bergin, E. A., et al. 2012, \aap, 542, 8
\bibitem[Lee et al.(2017)]{lee17} Lee, C.-F., Li, Z.-Y., Ho, P.~T.~P., et al.\ 2017, \apj, 843, 27
\bibitem[Lee et al.(2019)]{lee19} Lee, C.-F., Kwon, W., Jhan, K.-S., et al.\ 2019, arXiv e-prints, arXiv:1905.09417
\bibitem[Liseau et al.(1988)]{lis88} Liseau, R., Sandell, G., \& Knee, L. B. G. 1988, \aap, 192, 153
\bibitem[Liu et al.(2016)]{liu16} Liu, H. B., Lai, S.-P., Hasegawa, Y., et al. 2016, \apj, 821, 41
\bibitem[Liu et al.(2019)]{liu19} Liu, S.-Y. et al. 2019, in prep.
\bibitem[Looney, Mundy \& Welch(2000)]{loo00} Looney, L. W., Mundy, L. G., \& Welch, W. J. 2000, \apj, 529, 477
\bibitem[L\'{o}pez-Sepulcre et al.(2017)]{lop17} L\'{o}pez-Sepulcre, A., Sakai, N., Neri, R. et al. 2017, \aap, 606, A121
\bibitem[Mardones et al.(1997)]{mar97} Mardones, D., Myers, P. C., Tafalla, M., et al. 1997, \apj, 489, 719
\bibitem[Maret et al.(2004)]{mar04} Maret, S., Ceccarelli, C., Caux, E., et al. 2004, \aap, 416, 577
\bibitem[Mottram et al.(2013)]{mot13} Mottram, J. C., van Dishoeck, E. F., Schmalzl, M., et al. 2013, \aap, 558, 126
\bibitem[Myers, Evans, \& Ohashi(2000)]{mye00} Myers, P. C., Evans, N. J., II, \& Ohashi, N. 2000, Protostars and Planets IV, ed. V. Mannings, A. P. Boss, \& S. S. Russell (Tucson, AZ: Univ. Arizona Press), 217
\bibitem[Ohashi et al.(1997)]{oha97} Ohashi, N., Hayashi, M., Ho, P.~T.~P., et al.\ 1997, \apj, 475, 211
\bibitem[Ortiz-Le\'{o}n et al.(2018)]{ort18} Ortiz-Le\'{o}n G. N., Loinard L., Dzib S. A. et al 2018, \apj, 865, 73
\bibitem[Persson et al.(2016)]{per16} Persson, M. V., Harsono, D., Tobin, J. J. et al. 2016, \aap, 590, A33 
\bibitem[Sahu et al.(2019)]{dip19} Sahu, D., Liu, S.-Y., Su, Y.-N., et al. 2019, ApJ, in press
\bibitem[Sandell  \&  Knee(2001)]{san01} Sandell G., \& and Knee L. B. G. 2001, \apjl, 546, L49 
\bibitem[Sakai et al.(2014)]{sak14} Sakai, N., Sakai, T., Hirota, T., et al.\ 2014, \nat, 507, 78
\bibitem[Santangelo et al.(2015)]{san15} Santangelo, G., Codella, C., Cabrit, S., et al. 2015, \aap, 584, A126
\bibitem[Santiago-Garc\'{i}a et al.(2009)]{san09} Santiago-Garc\'{i}a, J., Tafalla, M., Johnstone, D., \& Bachiller, R. 2009, \aap, 495, 169
\bibitem[Segura-Cox et al.(2018)]{seg18} Segura-Cox, D. M., Looney, L. W., Tobin, J. J., et al. 2018, \apj, 866, 161
\bibitem[Shirley (2015)]{shi15} Shirley, Y. 2015, \pasp, 127, 299
\bibitem[Taquet et al.(2015)]{taq15}  Taquet, V., L\'{o}pez-Sepulcre, A., Ceccarelli, C., et al. 2015, \apj, 804, 81
\bibitem[Turner (1991)]{tur91}  Turner, B. E. 1991, \apjs, 76, 617
\bibitem[Wyrowski et al.(2012)]{wyr12} Wyrowski, F., G\"{u}sten, R., Menten, K. M., et al. 2012, \aap, 542, L15
\bibitem[Y$\i$ld$\i$z et al.(2013)]{yil13}Y$\i$ld$\i$z, U. A.,Acharyya, K.,Goldsmith, P. F., et al. 2013, \aap, 558, A58
\bibitem[Zapata et al.(2013)]{zap13} Zapata., L. A., Loinard., L., Rodr\'{i}guez, L. F., et al., 2013, \apjl, 764, L14 
\bibitem[Zucker et al.(2018)]{zuc18} Zucker, C., Schlafly, E. F., Green, G. M., et al. 2018, \apj, 869, 83
\end{thebibliography}
\end{document}